\documentclass[twocolumn]{autart}

\usepackage{color}
\usepackage{natbib}

\usepackage{cite}
\usepackage{graphicx}
\usepackage{theorem}
\usepackage{amsmath}
\usepackage{amssymb}
\usepackage{amsfonts}
\usepackage{arydshln}
\usepackage{balance}
\usepackage{mathtools}
\usepackage[subrefformat=parens,labelformat=parens]{subcaption}
\usepackage{url}

\theoremstyle{plain}
\newtheorem{definition}{Definition}[section]
\newtheorem{theorem}{Theorem}[section]
\newtheorem{lemma}{Lemma}[section]

\newcommand{\posdef}{\succ}
\newcommand{\negdef}{\prec}
\newcommand{\possemidef}{\succcurlyeq}
\newcommand{\negsemidef}{\preccurlyeq}

\begin{document}

\begin{frontmatter}

\title{Sampled-Data Controller Synthesis using Dissipative Linear Periodic Jump-Flow Systems with Design Applications}
%\textcolor{red}{revise title:}Dissipativity of Linear Periodic Jump-Flow Systems with Application to Sampled-Data Controller Synthesis}
\author{L.M. Spin}\ead{l.m.spin@tue.nl} \qquad
\author{M.C.F. Donkers}\ead{m.c.f.donkers@tue.nl}
\address{Dept. Electrical Engineering, Eindhoven University of Technology, Netherlands}

\begin{abstract}
In this paper, we will propose linear-matrix-inequality-based techniques for the design of sampled-data controllers that render the closed-loop system dissipative with respect to \textcolor{blue}{quadratic supply functions}, which includes passivity and an upper-bound on the system's $\mathcal{H}_\infty$-norm as a special case. To arrive at these results, we model the sampled-data control system as a linear periodic jump-flow system, study dissipativity in terms of differential linear matrix inequalities (DLMIs) and then convert these DLMIs into a single linear matrix inequality. We will present three applications of these synthesis techniques: 1) passivity-based controller synthesis, as found in teleoperations, 2) input-output-response matching of a continuous-time filter with a discrete-time filter (by minimizing the $\mathcal{H}_{\infty}$-norm of a generalized plant) and 3) a sampled-data controller redesign problem, where the objective is to find the best sampled-data controller, in the $\mathcal{H}_{\infty}$-norm sense, for a given continuous-time controller. We will show that synthesising sampled-data controllers leads to better closed-loop system behaviour than using a Tustin discretization of a continuous-time controller.
\end{abstract}
\end{frontmatter}

\section{Introduction}\label{introduction}

The vast majority of controller implementations have shifted from analogue to digital controllers over the last century. This paradigm shift has let to research on systems that have mixed continuous/discrete behaviour because the processor in-the-loop experiences discrete-time while the dynamics of the to-be-controlled process experiences continuous-time \citep{rosenwasser2000computer,aastrom2011computer}. This research is motivated by performance loss, and possibly instability, when discretizing continuous-time controllers for implementation on micro-controllers. When the hybrid nature of such systems is taken into account explicitly, performance and stability can be quantified and guaranteed, respectively, see e.g., \citep{chen2012optimal,bamieh1991lifting,mirkin1997mixed,Khargonekar1991}.

Control systems, in which the controller is implemented digitally and is explicitly incorporated in the design process, are called sampled-data control systems (SDCS). SDCSs are comprised of a continuous-time plant and sampler, a discrete-time controller and a hold device in closed-loop. This allows for direct design of a discrete-time controller for a continuous-time plant \citep{chen2012optimal}. 
In the 90s, SDCS design was very successful due to a technique called `lifting', see, e.g., \citep{bamieh1991lifting,Khargonekar1991}. While lifting transforms a SDCS into an equivalent infinite-dimensional discrete-time system, it is not directly applicable for systems that have discrete-time inputs and outputs \citep{mirkin1997mixed}, thereby limiting its applicability. Alternatively, the SDCS can be modelled as a system with delayed measurements, where the (time-varying) delay is reset to zero at every sampling instant, see, e.g., \citep{fridman2014}, though some conservatism is introduced in the analysis as the integral terms in the Lyapunov-Krasovskii functional are typically approximated, see \citep{Hetel2017}. Finally, SDCSs can be modeled as linear periodic jump-flow systems (LPJF), which form a distinct class of hybrid dynamical systems \citep{goebel2012hybrid}, which does not suffer from the  aforementioned limitations.

In recent years, a lot of attention has been given to LPJF systems and advancements in this area have enabled designing controllers for LPJF systems using Riccati differential and difference equations, see, e.g., \citep{ichikawa2001linear,sivashankar1994characterization}, which are not trivial to solve. In \citep{8114196,POSSIERI2020108772}, these Riccati differential and difference equations appear directly in the solution to the optimal state-feedback control problems and a tractable solution for problem is obtained using the so-called Monodromy Riccati equation. 
Alternatively, the Riccati differential/difference equations can are relaxed into differential linear matrix inequalities (DLMIs), see  \citep{geromel2023differential,HOLICKI2019179}. A DLMI can be turned into numerically tractable linear matrix inequalities (LMIs) by approximating the unknown variable as piecewise linear or using a Taylor/Fourier series \citep{geromel2023differential} or using splines or a matrix sum-of-square (SOS) approach \citep{HOLICKI2019179}. For the particular case of SDCSs, it is shown that these DLMIs can be turned into one LMI, which reduces the computational complexity, see \citep{dreef2021h,heemels2012periodic}.

In this paper, we aim to generalize the DLMI-based results for SDCSs presented in \citep{dreef2021h,geromel2023differential,HOLICKI2019179} towards  dissipativity, which been introduced for linear time-invariant systems in \citep{willems1972dissipative} and for jump-flow systems in \citep{5717501}. We will present controller synthesis results that render the SDCS dissipative with respect to a quadratic storage function in terms of solving one LMI. We will also indicate how this result can be used for synthesis for passivity-based control and to minimize the $\mathcal{H}_\infty$-norm of the closed-loop system. Compared to \citep{dreef2021h}, the $\mathcal{H}_\infty$ SDCS synthesis results will allow for more general generalized plant formulations. Finally, we will present applications of the presented synthesis results to three classes of control design problems. The first application is the synthesis of passive controllers for teleoperations, the second application is a sampled-data filter-matching problem, where the objective is to match the response of a continuous-time filter with a discrete-time filter by minimizing the $\mathcal{H}_{\infty}$-norm of the difference between the sampled-data and the continuous-time filter, and the third application is a sampled-data  controller redesign problem. In this latter problem, the objective is to find the best sampled-data controller, in the $\mathcal{H}_{\infty}$-norm sense, for a given continuous-time controller. For all three problems, we obtain better closed-loop system behaviour than using a Tustin discretization of a continous-time controller.

%\textcolor{blue}{The paper is organized as follows. After introducing some notational conventions below, Section 2 will introduce the SDCS under consideration and model it as a LPJF system. The dissipativity analysis and the controller synthesis results will be presented in Section 3.} Subsequently, we will present and illustrate the three classes of design problems in Section 5. Lastly, conclusions will be drawn in Section 6.

\paragraph*{Nomenclature}
The sets $\mathbb{R}$, $\mathbb{R}_{\geqslant 0}$, $\mathbb{R}_{> 0}$ and  $\mathbb{N}$ denotes the real numbers, non-negative real numbers, positive real numbers and non-negative integers, respectively. The transpose of a matrix $A \in \mathbb{R}^{n \times m}$, is denoted by $A^\top \in \mathbb{R}^{m \times n}$. The notation $A \posdef \mathbf{0}$, $A\possemidef \mathbf{0}$, $A \negdef \mathbf{0}$, and $A \negsemidef \mathbf{0}$ denotes that $A=A^\top$ is positive definite, nonnegative definite, negative definite and nonpositive definite, respectively. We write matrices of the form ${\tiny \begin{bmatrix} A & B \\ B^\top & C \end{bmatrix}}$ as ${\tiny\begin{bmatrix} A & \star \\ B^\top & C \end{bmatrix}}$ and the zero matrix and identity matrix are denoted by $\mathbf{0}$ and $I$, respectively. For a vector $x\in\mathbb{R}^n$, we denote by $\| x \| :=\sqrt{x^\top x}$ its $2$-norm. For a signal $x:\mathbb{N} \rightarrow \mathbb{R}^n$, we denote by $\| x \|_{\ell_2} = \sqrt{\sum_{k=0}^\infty \| x[k] \|^2}$ its $\ell_2$-norm, and for a signal $x:\mathbb{R}_{\geq0} \rightarrow \mathbb{R}^n$, we denote by $\| x \|_{\mathcal{L}_2} = \sqrt{\int_{0}^\infty \| x(t) \|^2} \mathrm{d}t$ its $\mathcal{L}_2$-norm, provided that these quantities are finite. Finally, a signal $x(t)$ evaluated at $t = t_k \in \mathbb{R}_{\geqslant 0}$ with $k \in \mathbb{N}$ is denoted as $x[k] = x(t_k)$ and its left-limit in the time interval $(t_k, t_{k+1}]$ is denoted as $x(t_k^+) = \lim_{t\downarrow t_k} x(t)$.

\section{Generalized Plant Formulation}\label{genplant}
\begin{figure}[t]
\centering
\includegraphics[width=0.95\columnwidth]{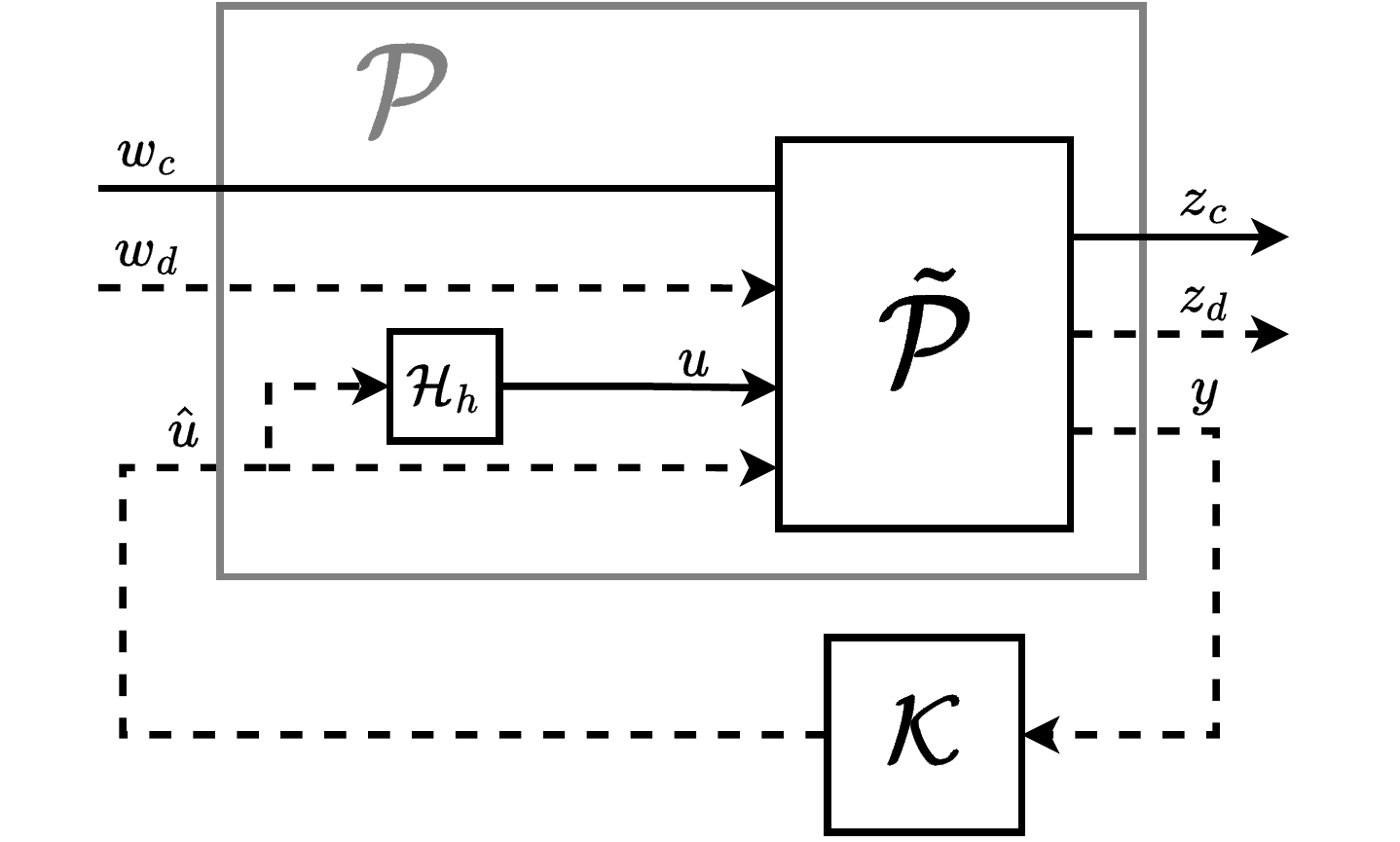}
\caption{The sampled-data control system schematic.}\label{fig:SD}
\end{figure}

In this paper, we will develop controller synthesis techniques for linear sampled-data control systems (SDCSs) with periodic sampling that render the SDCS dissipative with respect to a quadratic storage function. To do so, we model the SDCS as a so-called linear periodic jump-flow (LPJF) system, where the jump instances $t_k$ are periodic in time and are spaced by fixed $h\in\mathbb{R}_{> 0}$, i.e., $t_k = h k$, with jump index $k\in\mathbb{N}$. The SDCS is schematically shown in Fig.~\ref{fig:SD} and consists of a LPJF plant $\tilde{\mathcal{P}}$ connected with a hold device $\mathcal{H}_h$ that constrains the continuous-time (CT) input and the discrete-time (DT) input, leading to $\mathcal{P}=\tilde{\mathcal{P}}
\begin{bsmallmatrix}
I & 0 & 0  \\ 0 & I & 0 \\ 0 & 0 & \mathcal{H}_h\!\! \\ 0 & 0 & I 
\end{bsmallmatrix}
$, which is interconnected with a to-be-designed controller $\mathcal{K}$. It should be noted that $\hat{u}$ enters the plant $\tilde{\mathcal{P}}$ twice. Namely, it enters though the hold-device to act upon the physical system to-be-controlled, while it can also act directly on the weighting filters in $\tilde{\mathcal{P}}$ that are used to describe  performance requirements.

To develop a complete description of the sampled-data control problem, we assume the plant is given by 
\begin{equation}\label{eq:plantSD}
\tilde{\mathcal{P}} : \left\{\begin{array}{@{}r@{\ }l@{\ }c@{ }l@{\, }l@{\, }c@{ }l@{\, }c@{\, }c@{ }l}
\dot{x}(t)             &=& A_\mathrm{c} &x(t)   &+& B_\mathrm{c}   &w_\mathrm{c}(t) &+& B_\mathrm{c u} u(t)\\
x(t_k^+)               &=& A_\mathrm{d} &x(t_k) &+& B_\mathrm{d}   &w_\mathrm{d}[k] &+& B_\mathrm{d u} \hat{u}[k]\\
z_\mathrm{c}(t)        &=& C_\mathrm{c} &x(t)   &+& D_\mathrm{c c} &w_\mathrm{c}(t) &+& D_\mathrm{c u} u(t)\\
z_\mathrm{d}[k]        &=& C_\mathrm{d} &x(t_k) &+& D_\mathrm{d d} &w_\mathrm{d}[k] &+& D_\mathrm{d u} \hat{u}[k]\\
y[k]                   &=& C_\mathrm{y} &x(t_k) &+& D_\mathrm{y d} &w_\mathrm{d}[k] &+& D_{yu} \hat{u}[k]
\end{array}\right.
\end{equation}
for time $t\in(t_k,t_{k+1}]$, where $x(t) \in \mathbb{R}^{n_{\mathrm{\tilde{p}}}}$, $w_\mathrm{c}(t) \in \mathbb{R}^{n_{w_\mathrm{c}}}$, $w_\mathrm{d}[k] \in \mathbb{R}^{n_{w_\mathrm{d}}}$, $z_\mathrm{c}(t) \in \mathbb{R}^{n_{z_\mathrm{c}}}$, $z_\mathrm{d}[k] \in \mathbb{R}^{n_{z_\mathrm{d}}}$, $u(t) \in \mathbb{R}^{n_{u}}$, $\hat{u}[k] \in \mathbb{R}^{n_{u}}$, and $y[k] \in \mathbb{R}^{n_{y}}$ denote the state of the plant, the CT disturbance, the DT disturbance, the CT performance channels, the DT performance channel, the CT controller output, the DT controller output, the DT controller input, respectively. The CT input $u(t)$ and the DT input $\hat{u}[k]$ are constrained by a generalized hold device given by
\begin{equation}\label{eq:hold}
\mathcal{H}_h :
\left\{\begin{array}{@{}r@{\ }l@{\ }c@{ }l@{\, }l@{\, }c@{ }l@{\, }c@{\, }c@{ }l}
\dot{x}^\mathrm{h}(t) &=& A_\mathrm{c}^\mathrm{h} &x^\mathrm{h}(t)\\
x^\mathrm{h}(t_k^+)   &=& A_\mathrm{d}^\mathrm{h} &x^\mathrm{h}(t_k) &+& B^\mathrm{h} \hat{u}[k]\\
u(t)                  &=& C^\mathrm{h}            &x^\mathrm{h}(t).
\end{array}\right.
\end{equation}
with $x^\mathrm{h}(t)\in\mathbb{R}^{n_\mathrm{h}}$ denotes the state of the hold device. Choosing $A_\mathrm{c}^\mathrm{h}=A_\mathrm{d}^\mathrm{h}=\mathbf{0}$, $B^\mathrm{h}=C^\mathrm{h}=I$ corresponds to a traditional zero-order hold, although different and higher-order hold functions can be considered as well by possibly increasing the state dimension of the hold device $n_\mathrm{h}$. The plant and the hold device will be interconnected with a controller that has the following state-space realization:
\begin{equation}\label{eq:DiscreteControllerInJFFramework}
\mathcal{K}_\mathrm{d} : \left\{\begin{aligned}
\dot{x}^\mathrm{c}(t) &= \mathbf{0}\\
x^\mathrm{c}(t_k^+) &= A^\mathrm{c} x^\mathrm{c}(t_k) + B^\mathrm{c} y[k]\\
\hat{u}[k] &= C^\mathrm{c} x^\mathrm{c}(t_k) + D^\mathrm{c} y[k],
\end{aligned}\right.
\end{equation}
with controller state $x^\mathrm{c}(t) \in \mathbb{R}^{n_\mathrm{c}}$. The plant, hold device and controller are said to be in flow during the interval $t \in (t_k, t_{k+1}]$ and a jump occurs at $t=t_k=kh$. 

It should be noted that, besides considering more general hold devices in \eqref{eq:hold}, the plant description \eqref{eq:plantSD} is more general than the one used in \citep{dreef2021h}, where it was assumed that $B_\mathrm{d u}=\mathbf{0}$ and $D_\mathrm{c u}=\mathbf{0}$. Though this might seem a trivial extension, it will be shown in Section~\ref{sec:applications} that this extension is needed in order to apply the synthesis results that we will develop to the considered design problems. Furthermore, as was also argued in \citep{dreef2021h} and will be shown in Section~\ref{sec:applications}, it is important to consider jump dynamics in $\tilde{\mathcal{P}}$ to allow for DT weighing filters that express performance requirements. The only restriction that we place on $\tilde{\mathcal{P}}$ is that $D_\mathrm{yu}=0$, which may be done without loss of generality \citep{zhou1996robust}.

To arrive at a closed-loop description for the SDCS, we first combine the plant \eqref{eq:plantSD} with generalized hold device \eqref{eq:hold} to arrive at
\begin{equation}\label{eq:JFGenPlant}
\!\!\!\mathcal{P}:\left\{\begin{array}{@{}r@{\ }l@{\ }c@{ }l@{\, }l@{\, }c@{ }l@{\, }c@{\, }c@{ }l}
\dot{x}^\mathrm{p}(t)      &=& A_\mathrm{c}^\mathrm{p} &x^\mathrm{p}(t)   &+& B_\mathrm{c}^\mathrm{p}   &w_\mathrm{c}(t) & \\ 
x^\mathrm{p}(t_k^+)        &=& A_\mathrm{d}^\mathrm{p} &x^\mathrm{p}(t_k) &+& B_\mathrm{d}^\mathrm{p}   &w_\mathrm{d}[k] &+& B_\mathrm{J}^\mathrm{p}   &u_\mathrm{J}[k]\\
z_\mathrm{c}(t)            &=& C_\mathrm{c}^\mathrm{p} &x^\mathrm{p}(t)   &+& D_\mathrm{c c}^\mathrm{p} &w_\mathrm{c}(t) \\
z_\mathrm{d}[k]            &=& C_\mathrm{d}^\mathrm{p} &x^\mathrm{p}(t_k) &+& D_\mathrm{d d}^\mathrm{p} &w_\mathrm{d}[k] &+& D_\mathrm{d J}^\mathrm{p} &u_\mathrm{J}[k]\\
y[k]            &=& C_\mathrm{J}^\mathrm{p} &x^\mathrm{p}(t_k) &+& D_\mathrm{J d}^\mathrm{p} &w_\mathrm{d}[k],
\end{array}\right.
\end{equation}
where $x^\mathrm{p}(t) = [x(t)^\top \ x^\mathrm{h}(t)^\top]^\top \in \mathbb{R}^{n_\mathrm{p}}$, with matrices 
\begin{subequations}\label{eq:SDmatrices1}
\begin{align}
&A_\mathrm{c}^\mathrm{p} =\! \begin{bsmallmatrix}
\!A_\mathrm{c} & B_\mathrm{c u}C^\mathrm{h}\! \\ \mathbf{0}   & A_\mathrm{c}^\mathrm{h} \end{bsmallmatrix}, \quad\
B_\mathrm{c}^\mathrm{p} =\! \begin{bsmallmatrix} \!B_\mathrm{c u}\! \\ \mathbf{0} \end{bsmallmatrix}, \\%\quad\ B_\mathrm{F}^\mathrm{p}   = \mathbf{0}, \\ 
&A_\mathrm{d}^\mathrm{p} =\! \begin{bsmallmatrix} A_\mathrm{d} & \mathbf{0}  \\ \mathbf{0}   & A_\mathrm{d}^\mathrm{h}  \end{bsmallmatrix}, \qquad \
B_\mathrm{d}^\mathrm{p} =\! \begin{bsmallmatrix} B_\mathrm{d} \\ \mathbf{0} \end{bsmallmatrix}, \quad\ B_\mathrm{J}^\mathrm{p} =\! \begin{bsmallmatrix} \!B_\mathrm{d u}\!    \\ B^\mathrm{h} \end{bsmallmatrix}, \\
&C_\mathrm{c}^\mathrm{p}   =\! \begin{bsmallmatrix} C_\mathrm{c} & D_\mathrm{c u}C^\mathrm{h}  \end{bsmallmatrix}, \quad 
D_\mathrm{c c}^\mathrm{p} = D_\mathrm{c c}, \\%\quad\ D_\mathrm{c F}^\mathrm{p} = \mathbf{0}, \\
&C_\mathrm{d}^\mathrm{p}   =\! \begin{bsmallmatrix} C_\mathrm{d} & \mathbf{0} \end{bsmallmatrix}, \qquad\quad\!
D_\mathrm{d d}^\mathrm{p} = D_\mathrm{d d}, \quad
D_\mathrm{d J}^\mathrm{p} = D_\mathrm{d u}   \\
&C_\mathrm{J}^\mathrm{p} =\! \begin{bsmallmatrix} C_\mathrm{y} & \mathbf{0}  \end{bsmallmatrix}, \qquad\quad
D_\mathrm{J d}^\mathrm{p} = D_\mathrm{y d}, %\quad D_\mathrm{J J}^\mathrm{p} = \mathbf{0}.
\end{align}
\end{subequations}
Subsequently, we combine \eqref{eq:JFGenPlant} with \eqref{eq:DiscreteControllerInJFFramework}, with state $\xi(t) = [x^\mathrm{p}(t)^\top \ x^c(t)^\top]^\top\in\mathbb{R}^{n_\mathrm{p}+n_\mathrm{c}}$ to arrive at the LPJF system that is given by
\begin{equation}\label{eq:ClosedLoopSD}
\mathcal{T} : \left\{\begin{array}{@{}r@{\ }l@{\ }c@{ }l@{\, }l@{\, }c@{ }l@{\, }c@{\, }c@{ }l}
\dot{\xi} (t  ) &=& \mathcal{A}_\mathrm{c} &\xi(t)   &+& \mathcal{B}_\mathrm{c} &w_\mathrm{c}(t)\\
\xi(t_k^+)      &=& \mathcal{A}_\mathrm{d} &\xi(t_k) &+& \mathcal{B}_\mathrm{d} &w_\mathrm{d}[k]\\
z_\mathrm{c}(t) &=& \mathcal{C}_\mathrm{c} &\xi(t)   &+& \mathcal{D}_\mathrm{c} &w_\mathrm{c}(t)\\
z_\mathrm{d}[k] &=& \mathcal{C}_\mathrm{d} &\xi(t_k) &+& \mathcal{D}_\mathrm{d} &w_\mathrm{d}[k],
\end{array}\right.
\end{equation}
with flow matrices
\begin{subequations}\label{eq:SDmatrices}
  \begin{equation}\label{eq:SDflowmatrices}
\left[\begin{array}{c:c} \mathcal{A}_\mathrm{c} & \mathcal{B}_\mathrm{c} \\ \hdashline \mathcal{C}_\mathrm{c} & \mathcal{D}_\mathrm{c} \end{array}\right] = \left[ \begin{array}{cc:c} A    ^\mathrm{p}_\mathrm{c} & \mathbf{0} & B^\mathrm{p}_\mathrm{c} \\ \mathbf{0} & \mathbf{0} & \mathbf{0} \\ \hdashline C^\mathrm{p}_\mathrm{c} & \mathbf{0} & D^\mathrm{p}_\mathrm{c c} \end{array} \right]
\end{equation}
and the jump matrices 
\begin{equation}
\left[ \begin{array}{@{}c@{\, }:@{}c@{}} \!\mathcal{A}_\mathrm{d}\! & \ \mathcal{B}_\mathrm{d}\! \\ \hdashline \mathcal{C}_\mathrm{d} & \ \mathcal{D}_\mathrm{d} \!\end{array}\right] \!\!=\!\! \left[ \begin{array}{@{}c@{\ \ }c:c@{}} A^\mathrm{p}_\mathrm{d} & \mathbf{0} & \!\!B^\mathrm{p}_\mathrm{d} \\ \mathbf{0} & \mathbf{0} & \!\!\mathbf{0} \\ \hdashline C^\mathrm{p}_\mathrm{d} & \mathbf{0} & \! D^\mathrm{p}_\mathrm{d d} \end{array} \right] \!\!+\!\! \left[ \begin{array}{@{}c@{\ \,}c@{}} \mathbf{0} & B^\mathrm{p}_\mathrm{u} \\ I & \mathbf{0} \\ \hdashline \mathbf{0} & D^\mathrm{p}_\mathrm{d u} \end{array} \right] \!\!\!\begin{bmatrix} A^\mathrm{c} \!&\! B^\mathrm{c} \\ C^\mathrm{c} \!&\! D^\mathrm{c} \end{bmatrix}\!\!\!\left[ \begin{array}{@{}c@{\ \, }c:c@{}} \mathbf{0} & I & \!\mathbf{0} \\ C^\mathrm{p}_\mathrm{y} & \mathbf{0} & \!D^\mathrm{p}_\mathrm{y d}\end{array}\right]\!\!. \label{eq:JumpMatricesPartitioning}
\end{equation}
\end{subequations}
For this description of the SDCS, analysis and controller synthesis results that will be developed to warrant dissipativity and stability of the closed-loop system.

%\textcolor{red}{Refer to {\citep{possieri2016structural}} for the necessary structural properties.}
% that will be presented in this paper, it is important to know whether a controller \eqref{eq:DiscreteControllerInJFFramework} exists for the plant \eqref{eq:JFGenPlant} that renders \eqref{eq:ClosedLoopSD}  The existence of a exponentially stabilizing controller for plant \eqref{eq:JFGenPlant} requires stabilizabily and detectability, as studied in \citep{possieri2016structural}, which can be verified using the result in the following theorem. 
% \begin{theorem}\label{cor:genplant}
% System \eqref{eq:JFGenPlant} with matrices \eqref{eq:SDmatrices1} is a generalized plant if and only if it is stabilizable and detectable, i.e, if and only if
% \begin{subequations}
% \begin{align}
% \mathrm{rank}\big(\!\begin{bmatrix} \Phi_h-\lambda I & B_\mathrm{J}^\mathrm{p} \end{bmatrix}\!\big) = n_\mathrm{p}, \\
% \mathrm{rank}\big(\!\begin{bmatrix} \Phi_h^{\top} - \lambda I & ( C_\mathrm{J}^\mathrm{p} e^{A_\mathrm{c}^\mathrm{p} h} )^{\!\top} \end{bmatrix}^{\!\top}\!\big) = n_{\mathrm{p}}.
% \end{align}
% \end{subequations}
% holds for all $\lambda \in \Lambda(\Phi_h) \backslash\mathbb{D}$, in which $\Phi_h = e^{A_\mathrm{c}^\mathrm{p} h} A_\mathrm{d}^\mathrm{p}$.
% \end{theorem}}

\section{Stability \& Dissipativity of SDCSs}\label{dissipativity}

In this section, we will present general and tractable results to analyse stabily and dissipativity of SDCSs. We will first define dissipativity for the LPJF system \eqref{eq:ClosedLoopSD}, which has been defined for linear time-invarient systems in \citep{willems1972dissipative} and for jump-flow systems in \citep{5717501}. We will then focus on quadratic storage and supply functions and show that using techniques from \citep{goebel2012hybrid,heemels2012periodic}, analysis conditions can be formulated in the form of differential linear matrix inequalities (DLMIs). These DLMIs can then be turned into tractable conditions in terms of a single Linear Matrix Inequality (LMI), which can be used to develop controller synthesis results for sampled-data plant \eqref{eq:plantSD} with generalized hold \eqref{eq:hold}. Finally, we will show how to use these results to synthesize controllers that warrant that the SDCS is passive or has a certain upper bound on the $\mathcal{H}_\infty$-norm. We start by defining dissipativity of the LPJF system.

\begin{definition}
The LPJF system \eqref{eq:ClosedLoopSD} is said to be \emph{dissipative} with respect to the flow supply function $s_\mathrm{c}:\mathbb{R}^{n_\mathrm{w_\mathrm{c}}} \times \mathbb{R}^{n_\mathrm{z_\mathrm{c}}} \rightarrow \mathbb{R}$ and jump supply function $s_\mathrm{d}:\mathbb{R}^{n_\mathrm{w_\mathrm{d}}} \times \mathbb{R}^{n_\mathrm{z_\mathrm{d}}} \rightarrow \mathbb{R}$, if there exists a storage function $S:\mathbb{R}^{n_\mathrm{p}+n_\mathrm{k}} \times \mathbb{R}_{\geqslant 0} \rightarrow \mathbb{R}$ such that
\begin{equation}\label{eq:JFDissipativity}
\resizebox{1\hsize}{!}{$\displaystyle
S(\xi(t_1),t_1)\!\leqslant\!S(\xi(t_0),t_0)\!+ \!\!\int_{t_0}^{t_1}\!\!\!\!s_\mathrm{c}(w_\mathrm{c}(t),\!z_\mathrm{c}(t)\!) \mathrm{d}t \!+ \!\!\!\sum_{k = N_0}^{N_1} \!\!s_\mathrm{d}(w_\mathrm{d}[k], z_\mathrm{d}[k]),$}
\end{equation}
hold for all $t_1 \geqslant t_0$ and all $\xi(t_1)$ and $\xi(t_0)$ and all (essentially) bounded signals $w_\mathrm{c}$, $w_\mathrm{d}$, $z_\mathrm{c}$ and $z_\mathrm{d}$. Here, $N_0 \in \mathbb{N}$ denotes the smallest possible upper bound of $N_0 \geqslant t_0/h$ and $N_1 \in \mathbb{N}$ denotes the largest strict lower bound of $N_1 < t_1/h$. The system is said to be \emph{conservative} with respect to $s_\mathrm{c}$ and $s_\mathrm{d}$ if there exists a storage function $S$ such that equation \eqref{eq:JFDissipativity} holds with equality.
\end{definition}

Whenever the storage function $S(\xi, t)$ is at least once differentiable over the interval $(t_k, t_{k+1}]$ for all $k \in \mathbb{N}$ and all $\xi\in\mathbb{R}^{n_\xi}$, the inequality \eqref{eq:JFDissipativity} is equivalent to
\begin{subequations}\label{eq:dissInequalities}\color{blue}
\begin{align}
&\tfrac{\mathrm{d}}{\mathrm{d} t} S(\xi(t), t) \leqslant s_\mathrm{c}(w_\mathrm{c}(t), z_\mathrm{c}(t)), \ \ \mathrm{for} \ t\in(t_k, t_{k+1}]\!\!\label{eq:flowDiss}\\
&S(\xi(t_k^+), t_k^+) - S(\xi(t_k), t_k)\leqslant  s_\mathrm{d}(w_\mathrm{d}[k], z_\mathrm{d}[k]), \label{eq:jumpDiss}
\end{align}
\end{subequations}
where \eqref{eq:flowDiss} will be referred to as the flow dissipation inequality and \eqref{eq:jumpDiss} will be referred to as the jump dissipation inequality.

\subsection{Quadratic Storage and Supply Functions}

We will now focus on quadratic storage and supply functions. As in \citep{dreef2021h,goebel2012hybrid,heemels2012periodic}, we will consider a periodic storage function of the form
\begin{equation}
V(\xi, t) = \xi^{\top}\!(t) P(\tau(t)) \xi(t).\label{eq:Lyapfunc}
\end{equation}
%in which, the matrix function $P(\tau(t))$ is real, periodic, symmetric, bounded, positive definite and differentiable over $t \in (t_k, t_{k+1}]$ for  $\xi \in \mathbb{R}^\mathrm{n_{\xi}}$ and all $k\in\mathbb{N}_{0}$. 
%with timer function $\tau(t)$ is introduced, as in \citep{dreef2021h}, to emphasize the periodicity of $P(\cdot)$. 
with timer function $\tau(t)$, as in \citep{dreef2021h}, being a left-continuous function that satisfies:
\begin{equation}
\tau(t) = \begin{cases}
h \quad &\mathrm{for} \ t = 0,\\
t- t_k &\mathrm{for} \ t \in (t_k, t_{k+1}].
\end{cases}\label{eq:timer}
\end{equation}
meaning that $\tau(t_k) = h$ and $\tau(t_k^+) = 0^+$ for every $k \in \mathbb{N}$. We will also consider quadratic supply functions of the form
\begin{subequations}\label{eq:QRSsupply}
\begin{align}
s_\mathrm{c}(w_\mathrm{c}, z_\mathrm{c}) &=
\begin{bmatrix}w_\mathrm{c} \\ z_\mathrm{c}\end{bmatrix}^{\!\!\top}\begin{bmatrix} Q_\mathrm{c} & S_\mathrm{c} \\ S_\mathrm{c}^{\top} & R_\mathrm{c}\end{bmatrix} \begin{bmatrix}w_\mathrm{c} \\ z_\mathrm{c}\end{bmatrix} \label{eq:QRSflowSupply},\\
s_\mathrm{d}(w_\mathrm{d}, z_\mathrm{d}) &=
\begin{bmatrix}w_\mathrm{d} \\ z_\mathrm{d}\end{bmatrix}^{\!\!\top}\begin{bmatrix} Q_\mathrm{d} & S_\mathrm{d} \\ S_\mathrm{d}^{\top} & R_\mathrm{d}\end{bmatrix} \begin{bmatrix}w_\mathrm{d} \\ z_\mathrm{d}\end{bmatrix}. \label{eq:QRSjumpSupply}
\end{align}
\end{subequations}
The supply functions are inspired by a combination of \citep{goebel2012hybrid} and \citep{agarwal2016dissipativity}. The difference being that here both the flow and jump supply functions are quadratic and of similar structure. 

We will apply the considered quadratic storage and supply functions to obtain conditions for dissipativity and global exponentially stability (GES) of the LPJF system \eqref{eq:ClosedLoopSD}. By GES, we mean that all solutions satisfy $\|\xi(t)\| \leqslant c e^{-\alpha t} \|\xi(0)\|$ \textcolor{blue}{for all $t\in\mathbb{R}_{\geqslant 0}$ and all $\xi(0)\in\mathbb{R}^{n_\mathrm{p}+n_\mathrm{c}}$ and for some $c >0$ and $\alpha>0$.} In the lemma that we will present below, we will also use
\begin{equation}
%\resizebox{1\hsize}{!}{$\displaystyle
\begin{bmatrix} M_1 & M_3 \\ M_3^{\top} & M_2\end{bmatrix} \!=\! %\begin{bmatrix} \mathbf{0} & I \\ \mathcal{C}_\mathrm{c} & \mathcal{D}_\mathrm{c} \end{bmatrix}^{\!\top}\!\!\begin{bmatrix}Q_\mathrm{c} & S_\mathrm{c} \\ S_\mathrm{c}^{\top} & R_\mathrm{c}\end{bmatrix} \begin{bmatrix} \mathbf{0} & I \\ \mathcal{C}_\mathrm{c} & \mathcal{D}_\mathrm{c} \end{bmatrix} \notag \\ &\ \ = %
\begin{bmatrix} \mathcal{C}_\mathrm{c}^{\top} R_\mathrm{c} \mathcal{C}_\mathrm{c} &\mathcal{C}_\mathrm{c}^{\top} S_\mathrm{c}^{\top} + \mathcal{C}_\mathrm{c}^{\top} R_\mathrm{c} \mathcal{D}_\mathrm{c} \\ \star& Q_\mathrm{c}\!+\!S_\mathrm{c} \mathcal{D}_\mathrm{c} \!+\! \mathcal{D}_\mathrm{c}^{\top} S_\mathrm{c}^{\top} \!+\! \mathcal{D}_\mathrm{c}^{\top} R_\mathrm{c} \mathcal{D}_\mathrm{c}\end{bmatrix}.
\end{equation}

The results of the lemma are DLMIs and can be seen as an extension of the results of \citep{dreef2021h,geromel2023differential} towards dissipativity.

\begin{lemma}\label{Lem:DissAndStab}
Consider the LPJF system \eqref{eq:ClosedLoopSD} and supply functions \eqref{eq:QRSsupply} that satisfy \textcolor{blue}{$R_\mathrm{c} \preceq 0$ and $R_\mathrm{d}\preceq0$.
Assume there exists a bounded matrix function $P(\tau)\succ0$ that is differentiable on the domain $\tau \in (0, h]$ that satisfies}
\begin{subequations}\label{eq:dissthm}
\begin{align}\label{eq:RicDifEqLyap}
\resizebox{0.85\hsize}{!}{$\displaystyle
\!\!\!\begin{bmatrix} \dot{P}(\tau)\!+\!\mathcal{A}_\mathrm{c}^{\top} P(\tau)\!+\!P(\tau) \mathcal{A}_\mathrm{c} & P(\tau) \mathcal{B}_\mathrm{c} \\ \mathcal{B}_\mathrm{c}^{\top} P(\tau) & \mathbf{0}\end{bmatrix} \negsemidef \begin{bmatrix} M_1 & M_3 \\ M_3^{\top} & M_2 \end{bmatrix}\!\!,$}\!\!
\end{align}
for all $\tau \in (0, h]$ and
\begin{equation}    \label{eq:JumpLyapStabLMI}
\resizebox{0.95\hsize}{!}{$\displaystyle
\!\!\!\begin{bmatrix}\!\mathcal{A}_\mathrm{d}^{\top}\!P(0^+\!)\mathcal{A}_\mathrm{d} \!-\! P(h) & \mathcal{A}_\mathrm{d}^{\top}\!P(0^+\!) \mathcal{B}_\mathrm{d}\!\\ \mathcal{B}_\mathrm{d}^{\top}\!P(0^+\!) \mathcal{A}_\mathrm{d} \!\!&\!\! \mathcal{B}_\mathrm{d}^{\top}\!P(0^+\!) \mathcal{B}_\mathrm{d}\!\end{bmatrix}
\!\!\negsemidef\!\! \begin{bmatrix}\mathbf{0} \!\!&\!\! I \\ \mathcal{C}_\mathrm{d} \!&\! \mathcal{D}_\mathrm{d}\!\end{bmatrix}^{\!\!\top} \!\!\begin{bmatrix} Q_\mathrm{d} \!&\! S_\mathrm{d} \\ S_\mathrm{d}^{\top} \!\!&\!\! R_\mathrm{d} \end{bmatrix} \!\!\begin{bmatrix}\mathbf{0} \!\!&\!\! I \\ \mathcal{C}_\mathrm{d} \!&\! \mathcal{D}_\mathrm{d}\!\end{bmatrix}\!\!,$}
\end{equation}
\end{subequations}
where one of the inequalities holds strictly. Then, the LPJF system \eqref{eq:ClosedLoopSD} is dissipative with respect to supply functions \eqref{eq:QRSsupply} and its origin is GES.
\end{lemma}

\begin{pf}
Dissipativity follows directly from pre- and postmultiplying the conditions \eqref{eq:dissthm} with $[ \xi^\top \ w_c^\top]^\top$ and $[ \xi^\top \ w_d^\top]^\top$, respectively, leading to \eqref{eq:dissInequalities} with $S(\xi,t)=V(\xi,t)$ with \eqref{eq:Lyapfunc} and $s_c$ and $s_d$ as in \eqref{eq:QRSsupply}.

The strictness of one of the flow and jump dissipation inequalities \eqref{eq:dissthm} together with \textcolor{blue}{$R_\mathrm{c} \preceq 0 ,R_\mathrm{d}\preceq0$} and the assumption that $P(\tau)\succ0$ yields that \eqref{eq:Lyapfunc} becomes a \textcolor{blue}{Lyapunov function for the LPJF system} \eqref{eq:ClosedLoopSD}. Namely, in this case, we have that 
\begin{equation}
V(\xi(t),t) \leqslant e^{-\alpha_d}  e^{-\alpha_c (t-t_k)} V(\xi(t_k),t_k)\ \ \mathrm{for} \ t\in(t_k, t_{k+1}]
\end{equation}
with $\alpha_c\in\mathbb{R}_{\geqslant0}$ and $\alpha_d\in\mathbb{R}_{\geqslant0}$, where at least one of these inequalities holds strictly, meaning that $V(\xi(t),t) \leqslant e^{-\alpha_d(k+1)}  e^{-\alpha_c t} V(\xi(0),0)$ for $t\in(t_k, t_{k+1}]$. Now since $k+1\leqslant t/h$, we have that $\alpha = \alpha_d t/h + \alpha_c t>0$, if either $\alpha_c>0$ or $\alpha_d>0$. Now together with the fact that $P(\tau)$ is bounded for all $\tau\in(0,h]$, we have that $V$ is an exponentially decreasing Lyapunov function for the system, which concludes the proof.~\hfill$\square$
 % {\color{red}
 % The proof is based on showing that the conditions in the theorem imply \eqref{eq:dissInequalities} to hold and that one of the supply functions is strictly negative, see \eqref{eq:supply-stable}. To do so, first substitute the quadratic supply \eqref{eq:QRSsupply} and storage functions \eqref{eq:Lyapfunc} with the timer \eqref{eq:timer}, realising this renders $V(\xi,t)$ periodic, into \eqref{eq:flowDiss}, leading to
 % \begin{equation}
 % \!\!\!\begin{bmatrix} \dot{P}(\tau) + \mathcal{A}_\mathrm{c}^{\top} P(\tau) + P(\tau) \mathcal{A}_\mathrm{c}\!\!\! & P(\tau) \mathcal{B}_\mathrm{c} \\ \mathcal{B}_\mathrm{c}^{\top} P(\tau) & \mathbf{0}\end{bmatrix} - \begin{bmatrix} M_1 & M_3 \\ M_3^{\top} & M_2 \end{bmatrix} \negsemidef \mathbf{0},\!
 % \end{equation}
 % for $\tau\in(0,h]$. Now applying a Schur complement to this expression, which requires $M_2\posdef \mathbf{0}$, yields that this expression is satisfied by the satisfaction of \eqref{eq:RicDifEqLyap}. Secondly, applying the same supply and storage function to \eqref{eq:jumpDiss}, leads to \eqref{eq:JumpLyapStabLMI} with a nonstrict inequality. Therefore, both conditions \eqref{eq:dissInequalities} are satisfied. Furhermore, the fact that \eqref{eq:JumpLyapStabLMI} holds strictly, implies that a storage function can be defined that renders \eqref{eq:supply-stable_jump} with strict inequality, thereby rendering \eqref{eq:ClosedLoopSD} GES.}
\end{pf}

To arrive at tractable conditions for dissipativity and GES of the LPJF system \eqref{eq:ClosedLoopSD}, we can directly rely on numerical methods to solve DLMIs. Instead, we follow the approach from \citep{heemels2012periodic,goebel2012hybrid} in this paper, which is based on explicitly solving \eqref{eq:RicDifEqLyap} with equality. This approach employs the Hamiltonian matrix
\begin{equation}
\!\!H = \begin{bmatrix} \mathcal{A}_\mathrm{c} - \mathcal{B}_\mathrm{c} M_2^{-1} M_3^{\top} & \mathcal{B}_\mathrm{c} M_2^{-1} \mathcal{B}_\mathrm{c}^{\top} \\ M_1 - M_3 M_2^{-1} M_3^{\top} & - (\mathcal{A}_\mathrm{c} - \mathcal{B}_\mathrm{c} M_2^{-1} M_3^{\top})^{\top}\end{bmatrix}\!\!.\!\label{eq:HamiltonianMatrix}
\end{equation}
and its matrix exponential, in which we use the same partitioning as was used for $H$, given by
\begin{equation}
    F(\tau) = e^{-H \tau} = \begin{bmatrix} F_{1 1} (\tau) & F_{1 2} (\tau) \\ F_{2 1} (\tau) & F_{2 2} (\tau) \end{bmatrix}.
    \label{eq:HamilExponentPartition}
\end{equation}
% \begin{assumption}\label{A:HamiltonianExponent}
% $F_{1 1} (t)$ invertible for all $t \in [0, h]$.
% \end{assumption}
%Assumption~\ref{A:HamiltonianExponent} is always satisfied for sufficient small $h$ since $F(\tau)$ is a continuous function and $F_{1 1}(0) = I$. 
It should be noted that this Hamiltonian also appears in the computation of the Monodromy Riccati equation in \citep{8114196,POSSIERI2020108772}, showing the similarities between the approaches. Furthermore, $M_2$ is required to be invertible, which is a slight relaxation compared to \eqref{eq:RicDifEqLyap}. Finally, we define $\hat{A} = F_{1 1}^{-1}(h)$ and $\hat{B}$ and $\hat{C}$ satisfying
\begin{equation}\label{eq:structure}
\hat{B} \hat{B}^{\!\top}\!=\! -F_{1 1}^{-1}(h) F_{1 2}(h), \quad \hat{C}^{\!\top}\!\hat{C} = F_{2 1}(h) F_{1 1}^{-1}(h).
\end{equation}
to establish a close relation to results derived using lifting, see \citep{heemels2015mathcal}. \textcolor{blue}{Furthermore, we decompose
\begin{equation}
	R_\mathrm{d} = - E_\mathrm{d}^{\top} \Pi_\mathrm{d}^{-1} E_\mathrm{d} \preceq 0,
\end{equation}
conformly with the decomposition introduced in \citep{veenman2015general} for quadratic performance. Here $E_\mathrm{d}$ is the zero matrix or a constant full row rank matrix. Using this notation, we can obtain the following theorem that provides one LMI which guarantees dissipativity and GES.
}

\begin{theorem}\label{THM:DissAnalysisLMI}
Consider the LPJF system \eqref{eq:ClosedLoopSD}, assume that $F_{1 1} (t)$ is invertible for all $t \in [0, h]$, and that the quadratic supply functions are given by \eqref{eq:QRSsupply}, \textcolor{blue}{such that $\Pi_\mathrm{d} \succ \mathbf{0}$ and $M_2\succ\mathbf{0}$. If there exists a matrix $P_h$ satisfying
%\begin{equation}
%\begin{bmatrix}
%P_h & \star & \star & \star & \star\\
 %   S_\mathrm{d} \mathcal{C}_\mathrm{d} & Q_\mathrm{d} \!+\! S_\mathrm{d} \mathcal{D}_\mathrm{d} \!+\! \mathcal{D}_\mathrm{d}^{\top} S_\mathrm{d}^{\top} & \star & \star & \star\\ 
  %  \hat{C} \mathcal{A}_\mathrm{d} & \hat{C} \mathcal{B}_\mathrm{d} & I & \star & \star\\
   % P_h \hat{A} \mathcal{A}_\mathrm{d} & P_h \hat{A} \mathcal{B}_\mathrm{d} & \mathbf{0} & P_h & \star\\
   % \mathbf{0} & \mathbf{0} & \mathbf{0} & \hat{B}^{\top}\!P_h & I 
%\end{bmatrix}
%\!\!\posdef\!\mathbf{0}\!\label{eq:analLMIR0}
%\end{equation}
%whenever $R_\mathrm{d} = \mathbf{0}$, or satisfying
\begin{equation}
\!\!\begin{bmatrix}
    P_h & \star & \star & \star & \star & \star \\
    S_\mathrm{d} \mathcal{C}_\mathrm{d} & \!Q_\mathrm{d} \!+\! S_\mathrm{d} \mathcal{D}_\mathrm{d} \!+\! \mathcal{D}_\mathrm{d}^{\top} S_\mathrm{d}^{\top}\! & \star & \star & \star & \star \\ 
    E_\mathrm{d} \mathcal{C}_\mathrm{d} & E_\mathrm{d} \mathcal{D}_\mathrm{d} & \! \Pi_\mathrm{d} & \star & \star & \star \\
    \hat{C} \mathcal{A}_\mathrm{d} & \hat{C} \mathcal{B}_\mathrm{d} & \mathbf{0} & I & \star & \star \\
    \!P_h \hat{A} \mathcal{A}_\mathrm{d} & P_h \hat{A} \mathcal{B}_\mathrm{d} & \mathbf{0} & \mathbf{0} & P_h & \star \\ 
    \mathbf{0} & \mathbf{0} & \mathbf{0} & \mathbf{0} & \hat{B}^{\top}\! P_h & I \end{bmatrix}\!\!\posdef\!\mathbf{0}\!\! \label{eq:analLMI}
\end{equation}
then} the LPJF system \eqref{eq:ClosedLoopSD} is dissipative with respect to the quadratic supply functions \eqref{eq:QRSsupply} and its origin is GES.
\end{theorem}

\begin{pf}
We start the proof by applying a Schur complement on the bottom four rows and columns of \eqref{eq:analLMI}, leading to \eqref{eq:JumpLyapStabLMI},  where the inequality is strict and with $P_h\succ0$, and in which 
\begin{equation}\label{eq:ricatti}
P(h-\tau) = (F_{21}(\tau)+F_{22}(\tau) P_h ) (F_{11} (\tau) + F_{12}(\tau) P_h)^{-1}.
\end{equation}
with $P_h = P(h)$ for $\tau=0^+=0$, due to continuity of the matrix exponential. 

To complete the proof, it remains to show that satisfaction of \eqref{eq:ricatti} at $\tau=0$ is equivalent to satisfaction \eqref{eq:RicDifEqLyap} for all $\tau\in(0,h]$ and that $P(\tau)\succ0$. First, observe that, using a Schur compliment, \eqref{eq:RicDifEqLyap} is implied by  
\begin{align}\label{eq:RicDifEqLyap1}
&\!\!\!\tfrac{\mathrm{d}}{\mathrm{d}\tau}P(\tau) \negsemidef - \mathcal{A}_\mathrm{c}^{\top} P(\tau) - P(\tau) \mathcal{A}_\mathrm{c} + M_1 \notag \\
&\qquad\quad\ - (P(\tau) \mathcal{B}_\mathrm{c} - M_3) M_2^{-1} (\mathcal{B}_\mathrm{c}^{\!\top}P(\tau)  - M_3^{\!\top}),
\end{align}
for all $\tau\in(0,h]$ and where we have to relax $M_2 \possemidef \mathbf{0}$ to $M_2 \posdef \mathbf{0}$. Now, using the arguments in \citep{heemels2012periodic}, we can obtain that \eqref{eq:ricatti} is a well-defined solution to \eqref{eq:RicDifEqLyap1} with equality, if $F_{1 1}(\tau)$ is  invertible for all $\tau \in (0, h]$ and if 
\begin{equation}    \label{eq:PwellDefined}
\begin{bmatrix} P_h^{-1} & \hat{B} \\ \hat{B}^{\top} & I \end{bmatrix} \posdef \mathbf{0}.
\end{equation}
holds, which follows directly from the hypothesis of the theorem.~\hfill$\square$
\end{pf}

\subsection{Controller Synthesis}

Let us now use the results of Theorem \ref{THM:DissAnalysisLMI} to develop controller synthesis results that render the SDCS \eqref{eq:ClosedLoopSD} with matrices \eqref{eq:SDmatrices} GES and dissipative with respect to supply functions \eqref{eq:QRSsupply}. This requires generalizing the results of \citep{dreef2021h}.

The synthesis result is based on the observation that the matrices $\hat{A}$, $\hat{B}$ and $\hat{C}$ that appear in Theorem \ref{THM:DissAnalysisLMI}, which result from \eqref{eq:HamilExponentPartition}, have the following structure:
\begin{equation}\label{eq:HamilMatricesPartitioning}
\hat{A} = \begin{bmatrix}\bar{A} & \mathbf{0} \\ \mathbf{0} & I\end{bmatrix},\quad
    \hat{B} = \begin{bmatrix}\bar{B} \\ \mathbf{0}\end{bmatrix},\quad
    \hat{C} = \begin{bmatrix}\bar{C} & \mathbf{0}\end{bmatrix}.
\end{equation}
with $\bar{A}\in\mathbb{R}^{n_\mathrm{p}\times n_\mathrm{p}}$ and $I$ is an identity matrix of size $n_\mathrm{k}$. From now on, it will be assumed that the state dimension of the controller is equal to the state dimension of the plant, i.e, $n_\mathrm{p}=n_\mathrm{k}$, which allows introducing the following partitioning of $P_h$ as well as a new matrix $T$:
\begin{equation}\label{eq:LyapPartitioning}
P_h = \begin{bmatrix}Y \!&\! V \\ V^{\!\top} \!&\! \hat{Y}\end{bmatrix}\!\!, \quad P_h^{-1} = \begin{bmatrix}X \!&\! U \\ U^{\!\top} \!&\! \hat{X}\end{bmatrix}\!\!, \quad T = \begin{bmatrix}Y \!&\! I \\ V^{\!\top} \!&\! \mathbf{0}\end{bmatrix}\!\!,
\end{equation}
with $\hat{Y} = V^\top (Y - X^{-1}) V$ and $\hat{X} = V^{-1} (Y X Y - Y) V^{-\top}$ for nonsingular matrices $U$ and $V$ satisfying ${UV^\top = I-  XY}$. \textcolor{blue}{Furthermore, note that $P_h^{-1} T = \begin{bmatrix*} I & X \\ \mathbf{0} & U^{\top}\end{bmatrix*}$.} These matrices allow applying the linearizing transformation of \citep{scherer1997multiobjective} on the results of Theorem 3.1 to obtain synthesis LMIs that render the SDCS dissipative and GES, as is done in the following theorem.

\begin{theorem}\label{THM:DissipativitySYnthesis}
Consider the sampled-data plant \eqref{eq:plantSD} with generalized hold \eqref{eq:hold}, assume that $F_{1 1} (t)$ invertible for all $t \in [0, h]$ and $M_2 \succ \mathbf{0}$ holds and that the quadratic supply functions are given by \eqref{eq:QRSsupply}, \textcolor{blue}{such that $\Pi_\mathrm{d} \succ \mathbf{0}$. Assume there exist matrices $\Gamma, \Theta, \Upsilon, \Omega, Y, X$ such that
%\begin{equation}\label{eq:synthesisLMI0}
%\begin{bmatrix}
 %   Y & \star &  \star & \star & \star & \star & \star\\
  %  I & X & \star & \star &  \star & \star & \star\\
   % S_\mathrm{d}\Omega_\mathrm{c} & S_\mathrm{d} K_2 & Z & \star & \star  & \star & \star\\
    %\bar{C} \Omega_\mathrm{a} & \bar{C} K_1 & \bar{C}\Omega_\mathrm{b} & I & \star & \star & \star\\
 %   L_1 & \Gamma & L_2 & \mathbf{0} &  Y & \star & \star\\
  %  \bar{A} \Omega_\mathrm{a} & \bar{A} K_1 & \bar{A} \Omega_\mathrm{b} & \mathbf{0} &  I & X & \star\\
   % \mathbf{0} & \mathbf{0} & \mathbf{0} & \mathbf{0} & \bar{B}^{\top}Y & \bar{B}^{\top}& I
    %\end{bmatrix}\succ \mathbf{0}
%\end{equation}
%whenever $R_\mathrm{d} = \mathbf{0}$, or such that
\begin{equation}\label{eq:synthesisLMI}
\begin{bmatrix}
    Y & \star & \star & \star & \star & \star & \star & \star\\
    I & X & \star & \star & \star & \star & \star & \star\\
    S_\mathrm{d} \Omega_\mathrm{c} & S_\mathrm{d}K_2 & Z & \star & \star & \star & \star & \star\\
    E_\mathrm{d} \Omega_\mathrm{c} & E_\mathrm{d} K_2 & E_\mathrm{d} \Omega_\mathrm{d} & \Pi_\mathrm{d} & \star & \star & \star & \star\\
    \bar{C} \Omega_\mathrm{a} & \bar{C} K_1 & \bar{C} \Omega_\mathrm{b} & \mathbf{0} & I & \star & \star & \star\\
    L_1 & \Gamma & Y L_2 & \mathbf{0} & \mathbf{0} & Y & \star & \star\\
    \bar{A} \Omega_\mathrm{a} & \bar{A} K_1 & \bar{A} \Omega_\mathrm{b} & \mathbf{0} & \mathbf{0} & I & X & \star\\
    \mathbf{0} & \mathbf{0} & \mathbf{0} & \mathbf{0} & \mathbf{0} & \bar{B}^{\top}Y & \bar{B}^{\top}& I
    \end{bmatrix}\succ \mathbf{0}
\end{equation}
in which}
\begin{subequations}
\begin{align}
\begin{bmatrix}
\Omega_\mathrm{a} & \Omega_\mathrm{b} \\ \Omega_\mathrm{c} & \Omega_\mathrm{d}
\end{bmatrix} &= \begin{bmatrix}
    A_\mathrm{d}^\mathrm{p} & B_\mathrm{d}^\mathrm{p} \\ C_\mathrm{d}^\mathrm{p} & D_\mathrm{d d}^\mathrm{p}
    \end{bmatrix} +\begin{bmatrix} B_\mathrm{u}^\mathrm{p} \\ D_\mathrm{d u}^\mathrm{p} \end{bmatrix}\Omega \begin{bmatrix} C_\mathrm{y}^\mathrm{p} & D_\mathrm{y d}^\mathrm{p}
 \end{bmatrix} \\
\begin{bmatrix}     K_1 \\ K_2
    \end{bmatrix} &= \begin{bmatrix}
    A_\mathrm{d}^\mathrm{p} & B_\mathrm{u}^\mathrm{p} \\ C_\mathrm{d}^\mathrm{p} & D_\mathrm{d u}^\mathrm{p}
    \end{bmatrix} \begin{bmatrix} X \\ \Upsilon \end{bmatrix}\\
\begin{bmatrix} L_1 & L_2 \end{bmatrix} &= \begin{bmatrix} Y & \Theta \end{bmatrix}
\begin{bmatrix} \bar{A} A_\mathrm{d}^\mathrm{p} & \bar{A} B_\mathrm{d}^\mathrm{p} \\ C_\mathrm{y}^\mathrm{p} & D_\mathrm{y d}^\mathrm{p}
\end{bmatrix} \\
Z &= Q_\mathrm{d} + S_\mathrm{d} \Omega_\mathrm{d} + \Omega_\mathrm{d}^{\top} S_\mathrm{d}^{\top}.
\end{align}
\end{subequations}
Then, the controller \eqref{eq:DiscreteControllerInJFFramework} with
\begin{equation}\label{eq:statespacecontrollermatrices}
\!
\begin{bmatrix} A^\mathrm{c} \!&\! B^\mathrm{c}\! \\ C^\mathrm{c} \!&\! D^\mathrm{c}\! \end{bmatrix} \!\!=\!\! \begin{bmatrix}V &\! Y\!\bar{A} B_\mathrm{u}^\mathrm{p} \\ \mathbf{0} &\! I\end{bmatrix}^{\!-1} \!\!\begin{bmatrix}\Gamma \!-\! Y\!\bar{A} A_\mathrm{d}^\mathrm{p} X \!& \Theta \\ \Upsilon \!& \Omega \end{bmatrix} \!\!\begin{bmatrix}U^{\top} \!&\! \mathbf{0} \\ C_\mathrm{y}^\mathrm{p}\!X \!&\! I\end{bmatrix}^{\!-1}\!\!\!\!,\!\!\!
\end{equation}
where $U$ and $V$ are nonsingular and chosen to satisfy $U V^{\top} = I - X Y$, renders the SDCS \eqref{eq:plantSD}, \eqref{eq:hold} and \eqref{eq:DiscreteControllerInJFFramework} dissipative with respect to supply functions \eqref{eq:QRSsupply} and its origin GES.
\end{theorem}

\begin{pf}
It can be verified that by applying a congruence transformation with $\mathrm{diag}(P_h^{-1} T, I, I, P_h^{-1} T, I)$ to the LMI \eqref{eq:analLMI} it is transformed into \eqref{eq:synthesisLMI}, where the partitioning of the matrices \eqref{eq:JumpMatricesPartitioning}, \eqref{eq:HamilMatricesPartitioning} and \eqref{eq:LyapPartitioning} is used. The transformation \eqref{eq:statespacecontrollermatrices}
%\eqref{eq:nonlintrafo} 
then yields a controller synthesis LMI.~\hfill$\square$
%The controller matrices are determined by solving \eqref{eq:nonlintrafo}
% \begin{equation}
% \!\!\begin{bmatrix}\Gamma & \Theta \\ \Upsilon & \Omega \end{bmatrix} \!\!=\!\! \begin{bmatrix}Y\!\bar{A}A_\mathrm{d}^\mathrm{p} X &\! \mathbf{0} \\ \mathbf{0} &\!\mathbf{0}\end{bmatrix} \!\!+\!\! \begin{bmatrix}V \!& Y\!\!\bar{A} B_\mathrm{u}^\mathrm{p}\! \\ \mathbf{0} \!&\! I \end{bmatrix} \!\!\begin{bmatrix} A^\mathrm{c} \!&\! B^\mathrm{c} \\ C^\mathrm{c} \!&\! D^\mathrm{c} \end{bmatrix}\!\! \begin{bmatrix}U^{\top} \!&\! \mathbf{0} \\ \!C_\mathrm{y}^\mathrm{p} X &\! I \end{bmatrix}\!\!.\!\!  \label{eq:nonlintrafo}
% \end{equation}
\end{pf}

\subsection{Special Cases: Passivity and the $\mathcal{H}_\infty$ Norm}\label{sec:special}

The results of Theorem 3.2 can be directly used for two specific cases of dissipativity, which are passivity and the $\mathcal{H}_\infty$-norm. The concept of passivity is popular in electronic circuits as well as teleoperations and haptics \citep{hannaford1989stability}. Passivity for CT and DT linear time-invariant systems states that $w_\mathrm{c}^\top(t) z_\mathrm{c}(t) \leqslant 0$ and $w_\mathrm{d}^\top[k] z_\mathrm{d}[k] \leqslant 0$, respectively, which requires that $n_\mathrm{w_c} = n_\mathrm{z_c}$ and $n_\mathrm{w_d} = n_\mathrm{z_d}$. Passivity represents a sign-preserving property of a system, and this can be analysed for LPJF systems, by considering the supply functions \eqref{eq:QRSsupply} \textcolor{blue}{with $S_\mathrm{c} = I, S_\mathrm{d} = I, Q_\mathrm{c} = \mathbf{0}, Q_\mathrm{d} = \mathbf{0}, R_\mathrm{c} = \mathbf{0}, E_\mathrm{d} = \mathbf{0},$ and $\Pi_\mathrm{d} = I$.}

Besides passivity, the dissipativity can also be used to minimize (an upper bound on) the $\mathcal{H}_\infty$-norm of the SDCS \eqref{eq:ClosedLoopSD}, which is an induced norm and has been defined in \citep{toivonen1997sampled,sivashankar1994characterization,dreef2021h}. 

\begin{definition}
For a stable LPJF system \eqref{eq:ClosedLoopSD} with $\xi_0 = 0$, the $\mathcal{H}_{\infty}$ norm is the $\mathcal{L}_2 \times \ell_2$-induced norm:
\begin{equation}
\|\mathcal{T}\|_{\mathcal{H}_{\infty}} = \sup_{0 \ne (w_\mathrm{c}, w_\mathrm{d}) \in \mathcal{L}_2 \times \ell_2}
\frac{\big( \|z_\mathrm{c}\|_{\mathcal{L}_{2}}^2 + \|z_\mathrm{d}\|_{\ell_{2}}^2 \big)^{\!\frac{1}{2}}}{\big( \|w_\mathrm{c}\|_{\mathcal{L}_{2}}^2 + \|w_\mathrm{d}\|_{\ell_{2}}^2 \big)^{\!\frac{1}{2}}}
\end{equation}
\end{definition}

From this definition, we can observe that for any upper bound $\gamma$ on the $\mathcal{H}_\infty$-norm, it holds that
\begin{equation}
\tfrac{1}{\gamma} \big( \|z_\mathrm{c}\|_{\mathcal{L}_{2}}^2 + \|z_\mathrm{d}\|_{\ell_{2}}^2 \big)
\leqslant \gamma \big( \|w_\mathrm{c}\|_{\mathcal{L}_{2}}^2 + \|w_\mathrm{d}\|_{\ell_{2}}^2 \big)
\end{equation}
for all $w_\mathrm{c}(t)$ and all $w_\mathrm{d}[k]$, which can be rewritten as \eqref{eq:JFDissipativity} with a positive storage function and supply functions \eqref{eq:QRSsupply} \textcolor{blue}{with $Q_\mathrm{c} = \gamma I, Q_\mathrm{d} = \gamma I$, $R_\mathrm{c} = -\tfrac{1}{\gamma} I, R_\mathrm{d} = -\tfrac{1}{\gamma} I$, $S_\mathrm{c} = 0,$ and $S_\mathrm{d} = 0$, meaning that} the condition in Theorem 3.1 and 3.2 that $M_2\succ\mathbf{0}$ becomes $\gamma^2 I - \mathcal{D}_c^\top \mathcal{D}_c\succ\mathbf{0}$. As $\gamma$ enters the matrices \eqref{eq:HamiltonianMatrix} and, hence, the matrices $\hat{A}$, $\hat{B}$ and $\hat{C}$ in a complex (non-convex) way, the lowest upper-bound on the $\mathcal{H}_\infty$-norm is computed by minimizing $\gamma$ using a line search/bisection, while checking the conditions of Theorem 3.1 or Theorem 3.2, respectively.

\section{Applications to Control Design}\label{sec:applications}
In this section, we will apply the SDCS synthesis results to three different design problems. We aim to emphasize the formulation of the control problem in the form of the sampled-data plant \eqref{eq:plantSD}, where we will highlight how a CT system-to-be-controlled is augmented with possibly DT components, such as weighing filters. We will first employ the passivity-based synthesis result for an application in teleoperations and will then continue with two examples that can be formulated as an $\mathcal{H}_\infty$-norm minimization problem. The first of these examples is a filter-matching problem, where the objective is to match the frequency response of the CT filter with a DT filter, and the second is a \emph{closed-loop} controller-matching problem, where the objective is to approximate the \emph{closed-loop} response of an a priori designed CT controller interconnected with a CT plant by a DT controller interconnected to the same CT plant. Both matching examples will be shown to outperform traditional discretization methods. All the examples have been implemented using the \textit{PeriodicJumpFlowSystemsToolbox}\footnote{This toolbox can be downloaded from github.com/LuukSpin/PeriodicJumpFlowSystemsToolbox}, although solutions using DLMIs of Lemma 3.2 can be obtained as well, albeit at a larger computational complexity as the differential terms in the DMLIs need to be approximated, see \citep{geromel2023differential}.

\subsection{Application to teleoperations}

\begin{figure}[t]
    \centering
    \includegraphics[trim={2.5cm 0cm 2.5cm 0cm},clip,width = 0.45\textwidth]{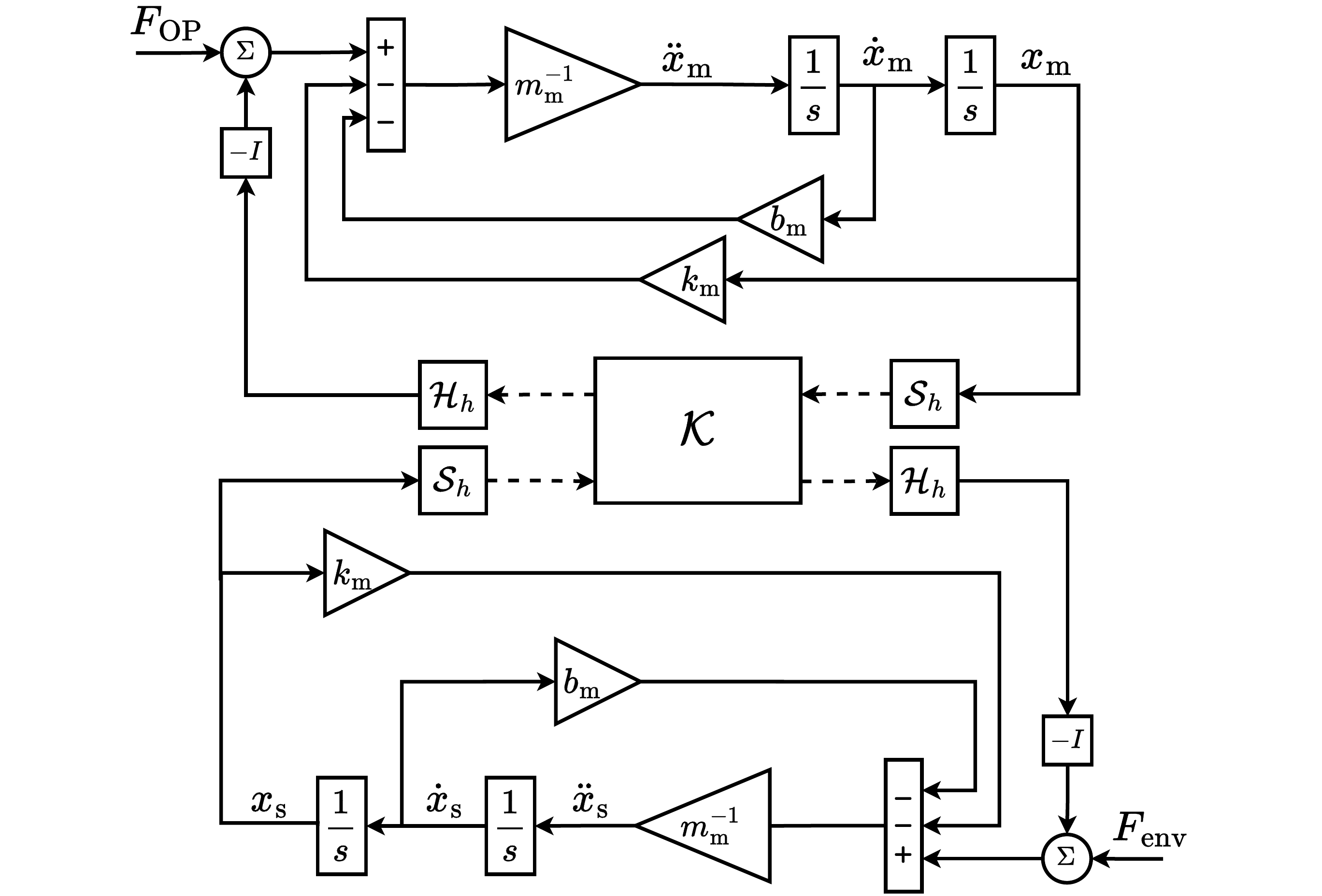}
    \caption{Passivity-based control of teleoperations scheme}
    \label{fig:passivitycontrolscheme}
\end{figure}

In teleoperations, the objective is to let a human operator control a master device and that behaviour is replicated (as closely as possible) by a slave device that interacts with an unknown environment which is assumed to be passive \citep{hogan1989controlling}. To let the human operator `feel' the unknown environment, the slave device has to feed the perceived force back to the master device. To achieve this, a controller interconnects the master and slave device. Since the human operator and the environment are uncertain, it is typically assumed that these are passive systems. An important property of passive systems is that interconnections of passive systems are passive \citep{van2000l2}. Therefore, if the master/slave system is passive, their interconnection with the human operator and the environment remains passive, and passivity can be used as a design tool in designing controller for teleoperation master/slave systems. In this example, we design a DT controller that controls the force of the CT master and CT slave system based on their position, see Fig.~\ref{fig:passivitycontrolscheme}.

In the teleoperation system shown in Fig.~\ref{fig:passivitycontrolscheme}, the master and slave devices are modelled using Newton's law with a certain mass $m_i$ and a viscous friction coefficent $b_i$, where $i\in\{m,s\}$ where $m$ and $s$ refer to the master and slave device, respectively. The controller that interconnects the physically disconnected systems measures the master and slave position and calculates a force, while both the operator and the environment also apply a force. This can be expressed using \eqref{eq:plantSD}, with
\begin{subequations}
\begin{align}
\!\!\!\!\!\!A_\mathrm{c} &=\!\! \begin{bsmallmatrix}
\!\!0 & 1 & 0 & 0 \\ \!\!\frac{-k_\mathrm{m}}{m_\mathrm{m}} & \frac{-b_\mathrm{m}}{m_\mathrm{m}} & 0 & 0 \\ 0 & 0 & 0 & 1 \\ 0 & 0 & \frac{-k_\mathrm{s}}{m_\mathrm{s}} & \frac{-b_\mathrm{s}}{m_\mathrm{s}} \!\! \end{bsmallmatrix}\!, \ 
B_\mathrm{c} = - B_\mathrm{cu}=\!\!\begin{bsmallmatrix} 0 & 0 \\
\!\!\frac{1}{m_\mathrm{m}} & 0 \\ 0 & 0 \\ 0 & \frac{1}{m_\mathrm{s}}\!\!
\end{bsmallmatrix}, \\
\!\!\!\!C_\mathrm{y} &=\! 
\begin{bmatrix}1 & 0 & 0 & 0 \\ 0 & 0 & 1 & 0\end{bmatrix}, \qquad \ D_\mathrm{y d} = \mathbf{0}, \qquad D_\mathrm{y u} = \mathbf{0}.
\end{align}
Indeed, these matrices express the two physically-disconnected master and slave device, where the operator and the environment apply a force, modelled through a vector $w_c(t)$. Since the system only evolves in CT, the DT dynamics of the plant are simply $
A_\mathrm{d} = I$, $B_\mathrm{d} = \mathbf{0}$ and $B_\mathrm{du} = \mathbf{0}$.
To express the requirement on passiviy, it should be noted that the product between input $w_c$ and output $z_c$ should represent a power, meaning that if the inputs are forces, the outputs have to be velocities. This is expressed by
\begin{equation}
C_\mathrm{c} =\! \begin{bsmallmatrix}\alpha k_\mathrm{m} & \alpha b_\mathrm{m} & -k_\mathrm{s} & - b_\mathrm{s} \\ 0 & 1 & 0 & -1\end{bsmallmatrix},  D_\mathrm{c c} = \begin{bsmallmatrix}\alpha & -1 \\ -\varepsilon & \varepsilon\end{bsmallmatrix},  D_\mathrm{c u} = \begin{bsmallmatrix} \alpha & -1 \\ 0 & 0 \end{bsmallmatrix},
\end{equation}
\end{subequations}
and $C_\mathrm{d} = \emptyset$, $D_\mathrm{d d} = \emptyset$ and $D_\mathrm{d u} = \emptyset$, where $\varepsilon$ is small in order to satisfy the constraint $M_2\succ\mathbf{0}$ and $\alpha$ is sufficiently large in order to guarantee that the force feedback to the operator is small. In this example, we take parameters $m_\mathrm{m} = 1$ [kg], $m_\mathrm{s} = 10$ [kg], $b_\mathrm{m} = 10^{-3}$ [Ns/m], $b_\mathrm{s} = 10^{-2}$ [Ns/m], $k_\mathrm{m} = 10$ [N/m], $k_\mathrm{s} = 10^{3}$ [N/m], $\epsilon=10^{-2}$ and $\alpha = 50$.

To arrive at a complete LPJF system representation of the SDCS given by \eqref{eq:JFGenPlant} with matrices \eqref{eq:SDmatrices}, we consider a zero-order hold (ZOH) and a first-order hold (FOH), which can be described by \eqref{eq:hold} with $A_\mathrm{c}^\mathrm{h}=A_\mathrm{d}^\mathrm{h}=\mathbf{0}$, $B^\mathrm{h}=C^\mathrm{h}=I$ for the ZOH and 
\begin{equation}\label{eq:FOH1}
A_\mathrm{c}^\mathrm{h} \!=\! \begin{bsmallmatrix}
\mathbf{0} & \mathbf{0} & \mathbf{0} \\
\!\frac{1}{h} I& \mathbf{0} & -\frac{1}{h} I\!\\
\mathbf{0} & \mathbf{0} & \mathbf{0}
\end{bsmallmatrix}\!, \ 
A_\mathrm{d}^\mathrm{h} \!=\! \begin{bsmallmatrix}
\mathbf{0} & \mathbf{0} & \mathbf{0} \\
 I& \mathbf{0} & \mathbf{0}\\
 I & \mathbf{0} & \mathbf{0}
\end{bsmallmatrix},\
B^\mathrm{h} \!=\! \begin{bsmallmatrix}
I \\ \mathbf{0} \\ \mathbf{0}
\end{bsmallmatrix}, \ 
C^\mathrm{h} \!=\! \begin{bsmallmatrix}
\mathbf{0} & I & \mathbf{0}
\end{bsmallmatrix}
\end{equation}
for the FOH. This SDCS description allows us to synthesize controller using the results of Theorem~\ref{THM:DissipativitySYnthesis}.

%Old text: \textcolor{red}{To compare the results, we compare the synthesis of DT controller with a CT controller that has been synthesized using the positive-real lemma, see \citep{scherer2000linear}, which is subsequently discretized using various discretization schemes and then analysed for passivity using the analysis results in Corollary~\ref{coro:PAss}. We now aim to maximize the sampling time $h$ for which these two methods can render the SD closed system passive (and GES). The obtained results are shown in Table \ref{table1}. It can be seen in this table, that the direct synthesis of DT controllers leads to a larger allowable sampling time $h$ than by discretizing a CT controller. In other words, the SD design approach results in a passive closed-loop system where the traditional design approach does not.}

Using the passivity synthesis technique given in Theorem~\ref{THM:DissipativitySYnthesis} \textcolor{blue}{with $S_\mathrm{c} = I, S_\mathrm{d} = I, Q_\mathrm{c} = \mathbf{0}, Q_\mathrm{d} = \mathbf{0}, R_\mathrm{c} = \mathbf{0}, E_\mathrm{d} = \mathbf{0}$, and $\Pi_\mathrm{d} = I$,} a DT controller is designed that renders the closed-loop GES and passive for both ZOH and FOH reconstructors irrespective of the sampling time. For example, for $h=10^{-3}$, a controller \eqref{eq:DiscreteControllerInJFFramework}, after making a balanced realization, \textcolor{blue}{is obtained with
\begin{equation}
\left[\begin{array}{@{}c:c@{}} A^c & B^c \\ \hdashline C^c & D^c \end{array}\right] \!\!=\! 
\resizebox{0.74\hsize}{!}{$\displaystyle
\left[ \begin{array}{@{}c@{\ }c@{\ }c@{ }c:c@{ }c@{}} 
0.979 & -0.12 & 1.15\!\cdot\!10^{-4} & 6.65\!\cdot\!10^{-4} & -1.34 & -6.95\\ 
0.12 & 0.16 & 0.068 & -0.032 & -0.3 & 20.65\\ 
-1.74\!\cdot\!10^{-3} & 0.023 & 0.995 & 4.9\!\cdot\!10^{-3} & 1.22 & -0.74 \\ 
1.9\!\cdot\!10^{-4} & -0.019 & -2.04\!\cdot\!10^{-3} & 0.02 
& 0.56 & 0.073%7.29\!\cdot\!10^{-2}
 \\ \hdashline
0.146 & 0.41 & -7.56\!\cdot\!10^{-3} & 0.018 & -1.13 & 59.9 \\ 
7.1 & 20.65 & -0.47 & 0.11 & -50.66 & 2936 
\end{array} \right]
$}
\end{equation}
% \begin{equation}
% A^c = \begin{bsmallmatrix} 0.9787 & -0.1179 & 1.145 \cdot 10^{-4} & 6.647 \cdot 10^{-4} \\ 0.1153 & 0.1557 & 0.06747 & -0.03187 \\ -1.744 \cdot 10^{-3} & 0.02318 & 0.9946 & 4. 875 \cdot 10^{-3} \\ 1.882 \cdot 10^{-4} & -0.01884 & -2.036 \cdot 10^{-3} & 0.01975  
% \end{bsmallmatrix}
% \end{equation}
% \begin{equation}
% B^c = \begin{bsmallmatrix} -1.344 & -6.95 \\ -0.2981 & 20.65 \\ 1.223 & -0.742 \\ 0.5603 & 0.07292 \end{bsmallmatrix}
% \end{equation}
% \begin{equation}
% C^c = \begin{bsmallmatrix} 0.1459 & 0.4104 & -7.562 \cdot 10^{-3} & 0.01748 \\ 7.096 & 20.65 & -0.4692 & 0.1132 \end{bsmallmatrix}
% \end{equation}
% \begin{equation}
% D^c = \begin{bsmallmatrix} -1.128 & 59.9 \\ -50.66 & 2936 \end{bsmallmatrix}
% \end{equation}
Alternatively,} a controller can be synthesized in continuous time using the positive-real lemma, see \citep{scherer2000linear}, leading to
\begin{equation}
\resizebox{1\hsize}{!}{$\displaystyle
\begin{cases}
\dot{x}^c(t) &=\!\begin{bmatrix} -2.89 & 0.983 & -16.4 & \!3.45\!\cdot\!10^{-3} \\ 4.12 \!\cdot\! 10^{4} & -401 & \!-8.25\!\cdot\!10^{4} & 61.4 \\ 2.18 & \!-8.55\!\cdot\!10^{-3} & -42.8 & 1 \\ 2.06\!\cdot\!10^{5} & -2\!\cdot\!10^3 & \!-4.09\!\cdot\!10^{5} & 307
\end{bmatrix} x^c(t) +\!\begin{bmatrix} -5.16 & -11.8 \\ -310 & 600 \\ 2.01 & -42.4 \\ -1925 & 5962 \end{bmatrix} y(t) \\
u(t) &= \begin{bmatrix} 4.12\!\cdot\!10^{4} & -402 & -8.30\!\cdot\!10^{4} & 61.4 \\ 2.06 \!\cdot\! 10^{6} & -2.01\!\cdot\!10^{4} & -4.15\!\cdot\!10^{6} & 3071 \end{bmatrix} x^c(t) + \begin{bmatrix} -352 & 213 \\ -1.76\!\cdot\!10^{4} & 1.06\!\cdot 10^{4} \end{bmatrix} y(t)
\end{cases}$}
\end{equation}
% \begin{equation}
% A^c = \begin{bsmallmatrix} -2.888 & 0.9829 & -16.36 & 3.453 \cdot 10^{-3} \\ 4.123 \cdot 10^{4} & -401.4 & -8.25 \cdot 10^{4} & 61.35 \\ 2.179 & -8.546 \cdot 10^{-3} & -42.75 & 1 \\ 2.056 \cdot 10^{5} & -2005 & -4.092 \cdot 10^{5} & 306.5
% \end{bsmallmatrix}
% \end{equation}
% \begin{equation}
% B^c = \begin{bsmallmatrix} -5.16 & -11.82 \\ -310.2 & 600 \\ 2.011 & -42.42 \\ -1925 & 5962 \end{bsmallmatrix}
% \end{equation}
% \begin{equation}
% C^c = \begin{bsmallmatrix} 4.124 \cdot 10^{4} & -401.7 & -8.296 \cdot 10^{4} & 61.4 \\ 2.062 \cdot 10^{6} & -2.009 \cdot 10^{4} & -4.148 \cdot 10^{6} & 3071 \end{bsmallmatrix}
% \end{equation}
% \begin{equation}
% D^c = \begin{bsmallmatrix} -351.8 & 212.8 \\ -1.759 \cdot 10^{4} & 1.063 \cdot 10^{4} \end{bsmallmatrix}
% \end{equation}
However, if this controller is subsequently discretized using various discretization schemes (e.g., ZOH, FOH or Tustin), and then analyzed for passivity using the results of Theorem~\ref{THM:DissAnalysisLMI}, the controller does not even render the SDCS stable. In other words, the SDCS design approach results in a passive closed-loop system whereas the traditional design approach does not.

%\begin{table}[t]
%\centering 
%\caption{Maximum sampling time $h$ for different discretization methods and hold devices used in the passivity analysis} \label{table1}
%\begin{tabular}{l|c|c}
 %                & \multicolumn{2}{c}{Analysis method} \\
%Synthesis method & ZOH  & FOH \\ \hline
%Direct \eqref{eq:PAssivitySyntheLMI} & 0.2275 & 0.2179 \\
%Tustin  & 0.1295 & 0.0651%09 
%\\
%FOH     & 0.1333 & 0.0668%1 
%\\
%ZOH     & 0.0460%4 
%& 0.0345%46 \\
%\end{tabular}
%\end{table}

\subsection{Application to Filter Matching}

\begin{figure}[t]
\centering
\includegraphics[trim={1.5cm 0cm 3cm 0cm},clip,width = \columnwidth]{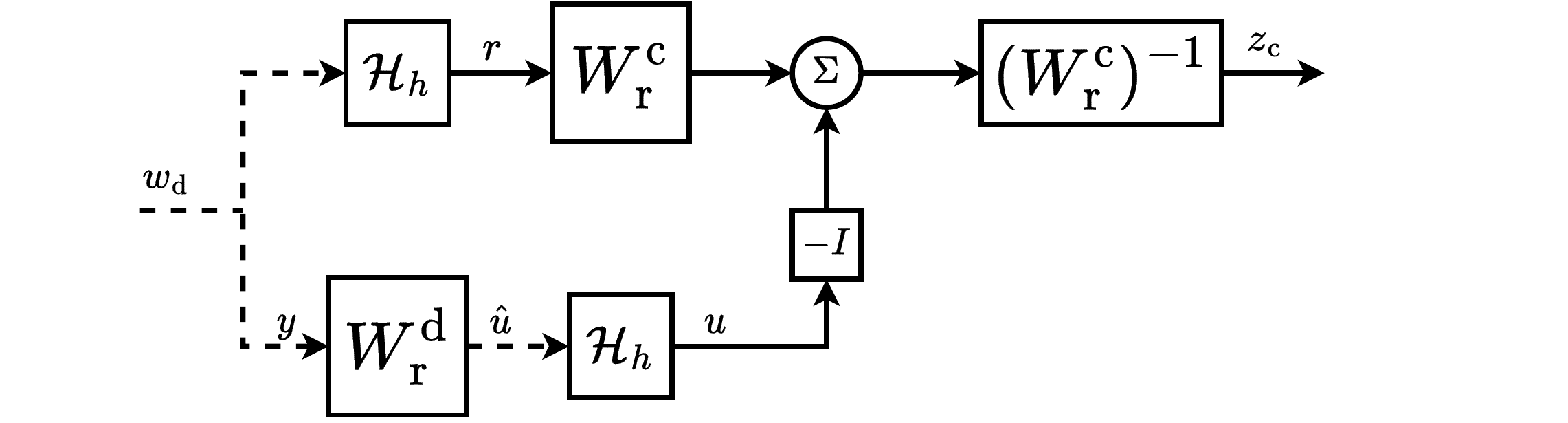}
\caption{Block diagram of the filter-matching problem.}
\label{fig:FilterMatching}
\end{figure}

The second application of the synthesis results presented in this paper is the problem of matching the response of a (stable) CT reference filter $W_\mathrm{r}^\mathrm{c}$ with a DT filter $W_\mathrm{r}^\mathrm{d}$. The block diagram of this problem is given in Fig.~\ref{fig:FilterMatching}. Here, matching is interpreted as minimizing the $\mathcal{H}_\infty$ norm of the (weighted) difference between the two filters, i.e., the difference from the DT input $w_d[k]$ to the (weighted) CT output $z_c(t)$. We apply the inverse of the CT reference filter $W_\mathrm{r}^\mathrm{c}$ to give every frequency equal emphasis in the synthesis problem. 

To obtain a sampled-data plant description of the form \eqref{eq:plantSD}, we regard $W_\mathrm{r}^\mathrm{d}$ as the to-be-designed controller, where the DT disturbance is directly fed to the output of the plant, i.e. $y[k] = w_d[k]$. Furthermore, the CT filter is assumed to have a state-space realization given by
\begin{subequations}
\begin{equation}
W_\mathrm{r}^\mathrm{c} : \left\{ \begin{aligned}
\dot{x}(t) &= A x(t) + B r(t)\\
x(t_k^+) &= I x(t_k)\\
z_\mathrm{r}(t) &= C x(t) + D r(t)
\end{aligned} \right.
\end{equation}
and to have a stable and causal inverse given by
\begin{equation}\label{eq:inverse}
\!\!\!(W_\mathrm{r}^\mathrm{c})^{-1}\!:\! \left\{ \begin{aligned}
   \dot{x}(t)   &= (A-B D^{-1} C) x(t) + B D^{-1} z_r(t)\\
    x(t_k^+)     &= I x(t_k)\\
    r(t) &= -D^{-1} C x(t) + D^{-1} z_r(t).
    \end{aligned} \right.\!
\end{equation}
It can be verified that the operator $(W_\mathrm{r}^\mathrm{c})^{-1} W_\mathrm{r}^\mathrm{c} = I$. In case $W_\mathrm{r}^\mathrm{c}$ does not have a stable and causal inverse, an approximate inverse $(W_\mathrm{r}^\mathrm{c})^\dagger$ needs to be found such that $(W_\mathrm{r}^\mathrm{c})^\dagger W_\mathrm{r}^\mathrm{c} \approx I$. Now the CT performance channel $z_c$ consists of the difference between $(W_\mathrm{r}^\mathrm{c})^{-1} W_\mathrm{r}^\mathrm{c} \mathcal{H}_h$ and the control input $u(t)$ that is filtered by $(W_\mathrm{r}^\mathrm{c})^{-1}$, i.e., $z_c(t) = \mathcal{H}_h w_d[k] - (W_\mathrm{r}^\mathrm{c})^{-1} u(t)$. The hold function here is taken as
\begin{align}\label{eq:hold2}
    \mathcal{H}_h &: \left\{ \begin{aligned}
    \dot{x}(t) &= \mathbf{0}\\
    x(t_k^+)   &= w_\mathrm{d}[k]\\
    r(t)       &= x(t).
    \end{aligned} \right.
\end{align}
\end{subequations}
Combining \eqref{eq:inverse} and \eqref{eq:hold2} with the fact that $y[k] = w_d[k]$ and $z_c(t) = \mathcal{H}_h w_d[k] - (W_\mathrm{r}^\mathrm{c})^{-1} u(t)$, leads to a plant description of the form \eqref{eq:plantSD} with
\begin{subequations}
\begin{align}
        A_\mathrm{c} &= \!\begin{bsmallmatrix}A - B D^{-1} C & \mathbf{0} \\ \mathbf{0} & \mathbf{0} \end{bsmallmatrix}, \ B_\mathrm{c} = \mathbf{0}, \quad\, B_\mathrm{u c} = \begin{bsmallmatrix} B D^{-1} \\ \mathbf{0} \end{bsmallmatrix},\\
        A_\mathrm{d} &= \!\begin{bsmallmatrix}I & \mathbf{0} \\ \mathbf{0} & \mathbf{0} \end{bsmallmatrix}, \qquad\qquad \, B_\mathrm{d} = \begin{bsmallmatrix}\mathbf{0} \\ I\end{bsmallmatrix}, \ B_\mathrm{u d} = \mathbf{0}\\
        C_\mathrm{c} &= \!\begin{bsmallmatrix}D^{-1} C & I \end{bsmallmatrix}, \qquad\, D_\mathrm{c c} = \mathbf{0}, \quad D_\mathrm{c u} = -D^{-1}\\
        C_\mathrm{d} &= \emptyset, \qquad\qquad\quad\, D_\mathrm{d d} = \emptyset, \quad D_\mathrm{d u} = \emptyset\\
        C_\mathrm{y} &= \mathbf{0}, \qquad\qquad\quad\ D_\mathrm{y d} = I, \quad\! D_\mathrm{y u} = \mathbf{0}
\end{align}
\end{subequations}
This plant description can be used to synthesize a controller, which in this case is the DT filter $W_\mathrm{r}^\mathrm{d}$, using Theorem~\ref{THM:DissipativitySYnthesis} and the procedure indicated in Section~\ref{sec:special}. It should be noted that the plant description above requires $D_\mathrm{c u} \neq \mathbf{0}$, meaning that the above problem could not have been solved using the plant description of \citep{dreef2021h}.

\begin{figure}[t]
\centering
\includegraphics[width = 0.9\columnwidth]{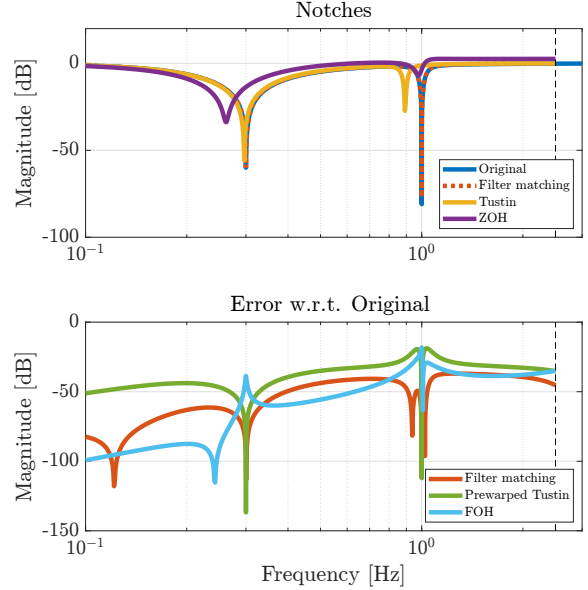}
\caption{Comparison of discretization techniques.}
\label{fig:FMAsDiscTechnique}
\end{figure}

To illustrate the filter-matching problem, let us consider a double notch filter described by the transfer function
\begin{equation}\label{eq:filtermatchgin}
    W_\mathrm{r}^\mathrm{c}(s) = \frac{\scalebox{0.95}{$s^4+0.00289 s^3 + 43.03 s^2 + 0.1 s + 140.3$}}{\scalebox{0.95}{$s^4+3.454 s^3 + 44.8 s^2 + 113.9 s + 140.3$}}
\end{equation}
and a sampling time of $h=0.2$ seconds. To assess the results, we compare the original CT filter with the result (after a balanced realization) of the above filter-matching problem, given by
\begin{equation}
\left[\begin{array}{@{}c:c@{}} A^c & B^c \\ \hdashline C^c & D^c \end{array}\right] \!\!=\! 
\resizebox{0.6\hsize}{!}{$\displaystyle
\left[ \begin{array}{@{}c@{\ }c@{\ }c@{ }c:c@{ }c@{}} 
0.236 & -0.307 & 0.474 & -0.177 & -0.567 \\  
0.307 & 0.924 & 0.120 & -0.0403 & -0.138 \\ 
-0.474 & 0.120 & 0.343 & -0.773 & 0.0913 \\ 
-0.177 & 0.0403 & 0.773 & 0.536 & 0.160 \\ \hdashline
0.567 & -0.138 & 0.0913 & -0.160 & 0.727 
\end{array} \right]
$}
\end{equation}
and several traditional discretization methods of \eqref{eq:filtermatchgin}, in terms of their Bode magnitude plot. The results are given in Fig.~\ref{fig:FMAsDiscTechnique}, where we have plotted the frequency responses of both CT and DT filters in a single figure and the error is computed pointwise-in-frequency. It can be seen from the top plot that the Bode magnitude of the original filter \eqref{eq:filtermatchgin} and the matched filter are very similar, while the ZOH and Tustin discretization of filter perform much worse. The bottom plot compares the matched filter with a FOH and a `Tustin with prewarp' discretization, but this time in terms of error between the DT filter and the CT reference filter, as the differences are small. Still, it can be concluded that, for this example, the SD-filter-matching approach presented here gives better approximations than traditional discretization methods.

% {\color{red} The obtained discrete-time filter $W_\mathrm{r}^\mathrm{d}$ using the $\mathcal{H}_{\infty}$ controller synthesis method has the following state-space matrices
% \begin{equation}
% A^c = \begin{bsmallmatrix} 0.2358 & -0.3066 & 0.4743 & -0.1766 \\ 0.3066 & 0.9239 & 0.1199 & -0.04025 \\ -0.4743 & 0.1199 & 0.3425 & -0.7725 \\ -0.1766 & 0.04025 & 0.7725 & 0.5355 \end{bsmallmatrix}
% \end{equation}
% \begin{equation}
% B^c = \begin{bsmallmatrix}  \end{bsmallmatrix}
% \end{equation}
% \begin{equation}
% C^c = \begin{bsmallmatrix} 0.5666 & -0.1377 & 0.09132 & -0.1596 \end{bsmallmatrix}
% \end{equation}
% \begin{equation}
% D^c = \begin{bsmallmatrix} 0.7266 \end{bsmallmatrix}
% \end{equation}
% }

\subsection{Application to Controller Matching}

Let us now consider the problem of designing a DT controller that matches the behaviour of a given CT controller. Here, it is assumed that the CT controller is designed using any design procedure (e.g., classical frequency response methods, or $\mathcal{H}_\infty$-based mixed sensitivity designs) such that desired closed-loop performance requirements are satisfied. Besides matching the input-output behaviour of the CT controller, the DT controller is also required to render the closed-loop SDCS stable, which is not guaranteed by directly discretizing the CT controller, using, e.g., a Tustin discretization.

\begin{figure}[t]
\centering
\includegraphics[trim={1cm 0cm 1.4cm 0cm},clip,width = \columnwidth]{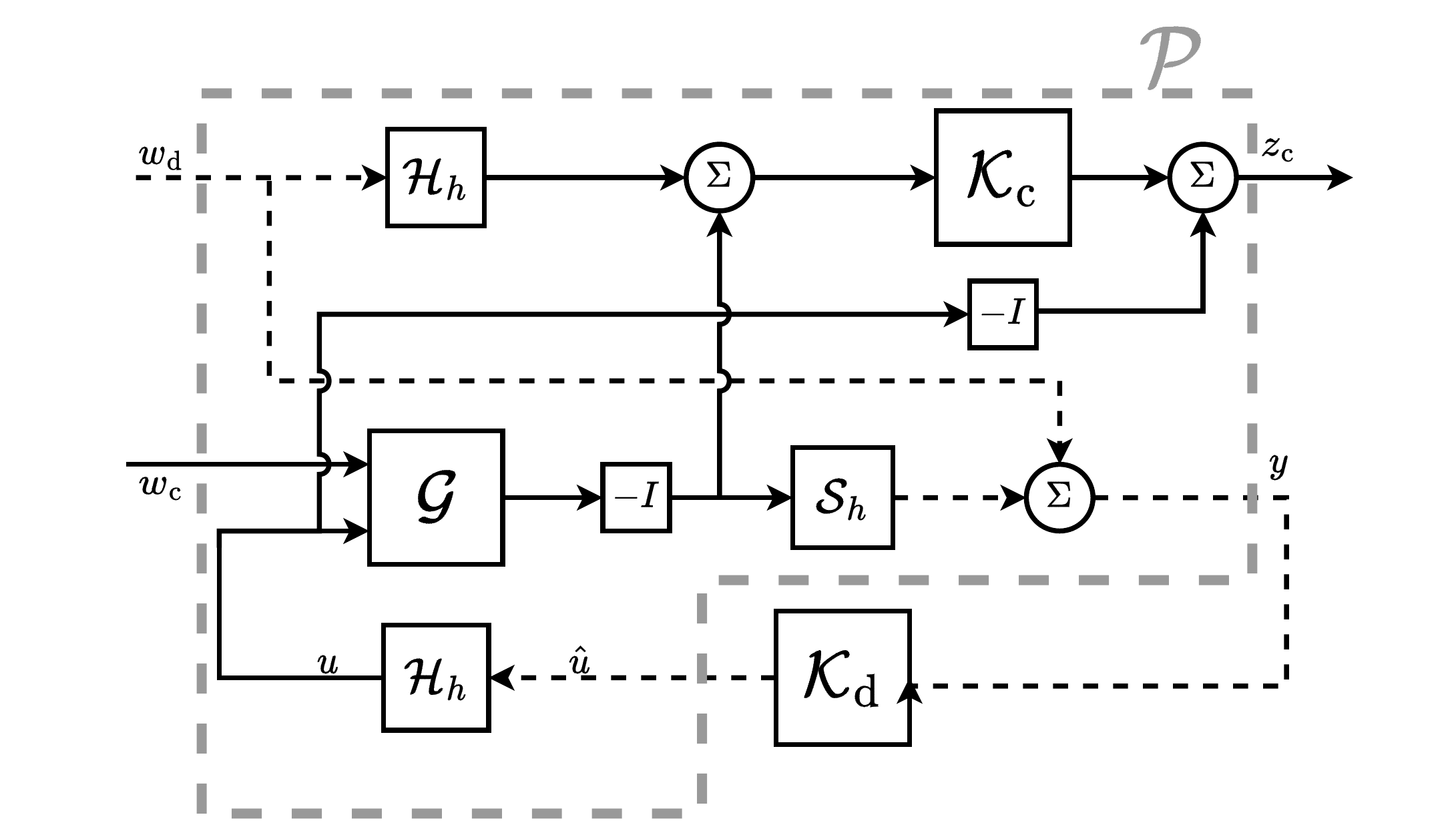}
\caption{Filter-matching technique for redesign of CT filters for SDCS synthesis.}\label{fig:FilterAndControllerMatching}
\end{figure}

The controller redesign problem is schematically shown in Fig.~\ref{fig:FilterAndControllerMatching}. In this figure, the system-to-be-controlled is assumed to be strictly proper and given by
\begin{subequations}\label{eq:redesign}
\begin{equation}
\mathcal{G} : \left\{ \begin{aligned}
\dot{x}^\mathrm{g}(t) &= A^\mathrm{g} x^\mathrm{g}(t) + B^\mathrm{g}_\mathrm{w} w_\mathrm{c}(t) + B^\mathrm{g}_\mathrm{u} u(t) \\
y_\mathrm{g}(t) &= C^\mathrm{g} x(t)
\end{aligned} \right.
\end{equation}
and the CT controller is given by
\begin{equation}
\!\!\!\mathcal{K}_\mathrm{c}\!:\! \left\{ \begin{aligned}
   \dot{x}^\mathrm{c}(t)   &= A^\mathrm{c} x^\mathrm{c}(t) + B^\mathrm{c} e(t)\\
    u(t) &= C^\mathrm{c} x^\mathrm{c}(t) + D^\mathrm{c} e(t).
    \end{aligned} \right.\!
\end{equation}
\end{subequations}
which is designed to ensure that \eqref{eq:redesign} with $e(t) = r(t) - y_\mathrm{g}(t)$ for a given CT reference $r(t)$ has desired closed-loop behaviour. In the redesign problem, a sampled-data controller $\mathcal{K}_\mathrm{d}$ is interconnected with $\mathcal{G}$ through a sampler and a ZOH, where the sampled-data controller takes $e[k] = w_\mathrm{d}[k] - \mathcal{H}_h y_\mathrm{g}(t)$ as an input, where $w_\mathrm{d}[k]$ is the DT reference. The objective is to minimize (in an $\mathcal{H}_\infty$-sense) the difference between output of $\mathcal{K}_\mathrm{c}$ and the ZOH-output of $\mathcal{K}_\mathrm{d}$ for (any) disturbance $w_\mathrm{c}$ acting on the to-be-controlled system $\mathcal{G}$ and (any) reference $w_\mathrm{d}$. This can be expressed in terms of operators
\begin{subequations}\label{eq:controllermatching}
\begin{align}
\!\!\!z_\mathrm{c}(t) &= \mathcal{K}_\mathrm{c} \left(\mathcal{H}_h w_\mathrm{d}[k]-\mathcal{G} \begin{bsmallmatrix}  w_\mathrm{c} (t)\\ u(t)\end{bsmallmatrix} \right) -u(t)\!
\end{align}
with plant output
\begin{equation}
y[k] = w_\mathrm{d}[k] - \mathcal{S}_h \mathcal{G} \begin{bsmallmatrix} w_\mathrm{c} (t)\\ u(t)\end{bsmallmatrix} 
\end{equation}
\end{subequations}
where $\mathcal{S}_h$ denotes a sampler, that samples the CT plant output at time instances $t_k$. 
To allow the synthesis problem to be solved, we define $x^\mathrm{p} = [ (x^\mathrm{g})^\top \  (x^\mathrm{h})^\top \  (x^\mathrm{c})^\top ]^\top$ as state, allowing \eqref{eq:controllermatching} to be written as
\begin{equation}
\left\{ \begin{array}{@{}r@{\,}l@{ }l@{ }l@{ }l@{}l@{}l@{}l@{}}
\dot{x}^\mathrm{p}(t) &=& \begin{bsmallmatrix} A^\mathrm{g} & 0 & 0 \\ 0 & 0 & 0 \\ -B^\mathrm{c} C^\mathrm{g} & B^\mathrm{c} & A^\mathrm{c} \end{bsmallmatrix} x^\mathrm{p}(t) +\!\! \begin{bsmallmatrix} B^\mathrm{g}_\mathrm{w} \\ 0 \\ 0 \end{bsmallmatrix} w_\mathrm{c}(t) + \!\begin{bsmallmatrix} B_\mathrm{u}^\mathrm{g} \\ 0 \\ 0 \end{bsmallmatrix} u(t) \\[3pt]
x^\mathrm{p}(t_k^+)   &=& \begin{bsmallmatrix} I   & 0 & 0 \\ 0 & 0 & 0 \\ 0        & 0   & I   \end{bsmallmatrix} x^\mathrm{p}(t_k) + \begin{bsmallmatrix} 0 \\ I \\ 0 \end{bsmallmatrix} w_\mathrm{d}[k] + \begin{bsmallmatrix} 0     \\ 0 \\ 0 \end{bsmallmatrix} \hat{u}(t) \\
z_\mathrm{c}(t) &=& \begin{bsmallmatrix} -D^\mathrm{c} C^\mathrm{g} & D^\mathrm{c} & C^\mathrm{c} \end{bsmallmatrix} x(t)  - u(t) \\ 
y[k]   &=& \begin{bsmallmatrix} -C^\mathrm{g}    & 0   & 0   \end{bsmallmatrix} x(t_k) + w_\mathrm{d}[k] 
\end{array} \right.
\end{equation}
%\begin{align*}
%        A_\mathrm{c} &= \begin{bmatrix}A_c & \mathbf{0} & \mathbf{0}\\ \mathbf{0} & \mathbf{0} & \mathbf{0} \\ B^c C_y & B^c & A^c \end{bmatrix}, \ B_\mathrm{c} = \begin{bmatrix} B_c \\ \mathbf{0} \\ \mathbf{0} \end{bmatrix}, \ B_\mathrm{u c} = \begin{bmatrix}B_{u c} \\ \mathbf{0} \\ \mathbf{0} \end{bmatrix}\\
%        A_\mathrm{d} &= \begin{bmatrix} I & \mathbf{0} & \mathbf{0} \\ \mathbf{0} & \mathbf{0} & \mathbf{0} \\ \mathbf{0} & \mathbf{0} & \mathbf{0} \end{bmatrix}, \ B_\mathrm{d} = \begin{bmatrix}\mathbf{0} \\ I \\ \mathbf{0} \end{bmatrix}, \ B_\mathrm{u d} = \begin{bmatrix}\mathbf{0} \\ \mathbf{0} \\ \mathbf{0}\end{bmatrix}\\
%        C_\mathrm{c} &= \begin{bmatrix}D^c C_y & D^c & C^c \end{bmatrix}, \ D_\mathrm{c c} = \mathbf{0}, \ D_\mathrm{c u} = -I\\
%        C_\mathrm{d} &= \emptyset, \ D_\mathrm{d d} = \emptyset, \ D_\mathrm{d u} = \emptyset\\
%        C_\mathrm{y} &= \begin{bmatrix}C_y & \mathbf{0} & \mathbf{0} \end{bmatrix}, \ D_\mathrm{y d} = D_\mathrm{y d}, \ D_\mathrm{y u} = D_\mathrm{y u}
%\end{align*}
which has the form of \eqref{eq:plantSD} which can be used to formulate \eqref{eq:JFGenPlant} whose matrices have the form \eqref{eq:SDmatrices}, which can be used in the results of Theorem~\ref{THM:DissipativitySYnthesis} and the procedure outlined in Section~\ref{sec:special}. 

\begin{figure}[t]
    \centering
    \includegraphics[width = \columnwidth]{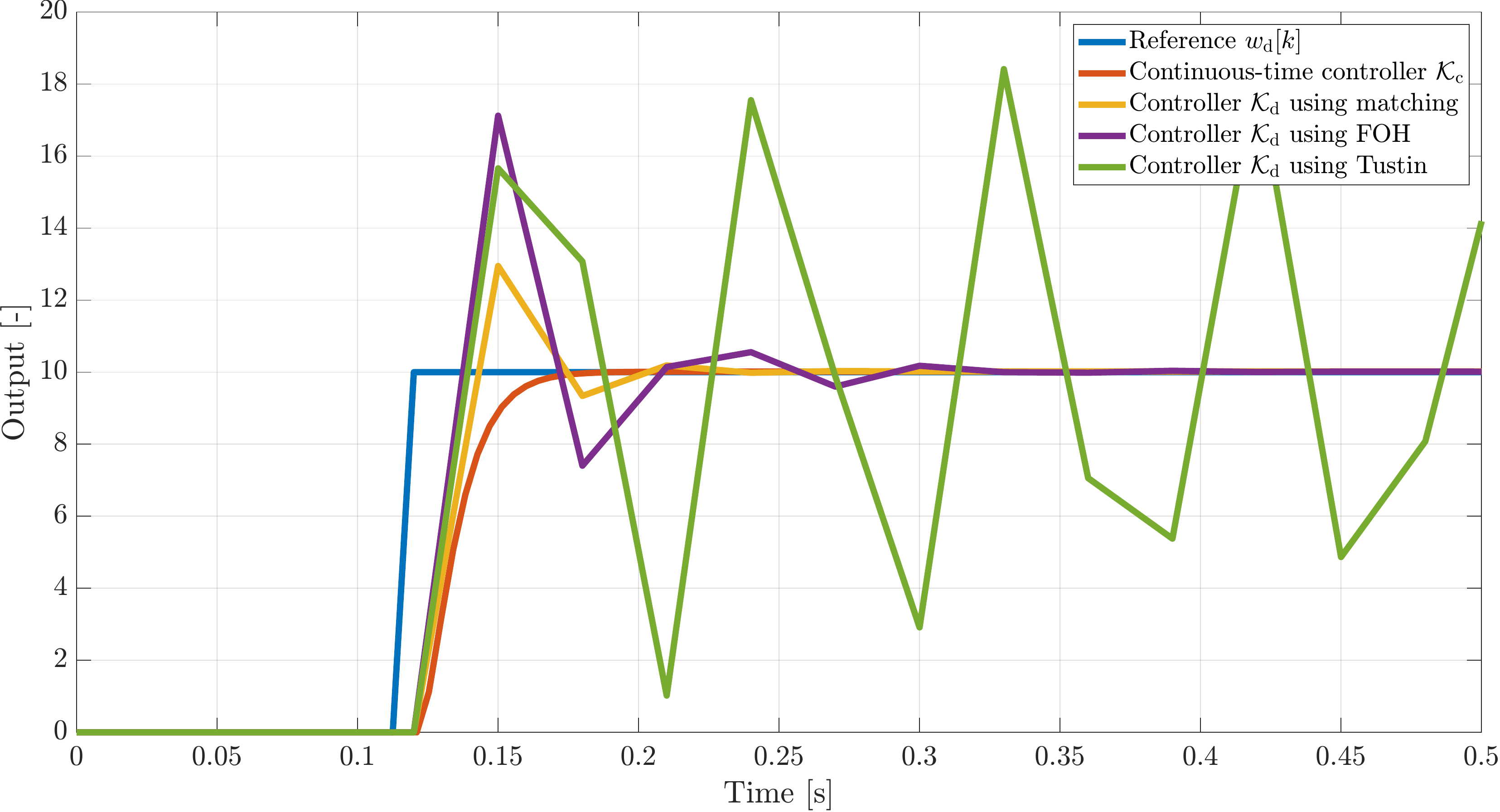}
    \caption{Controller-matching compared with discretization.}
    \label{fig:ControllerMatchingUnstableSys}
\end{figure}

To illustrate the controller redesign problem outlined above, let us consider a plant $\mathcal{G}$ and an a priori CT controller $\mathcal{K}_\mathrm{c}$, whose transfer functions satisfy
\begin{equation}
    G(s) = \frac{0.1}{s-0.1}. \qquad K_\mathrm{c}(s) = \frac{1.747\cdot10^5}{s+268.5}
\end{equation}
respectively. Note that in this example the CT disturbance $w_\mathrm{c}$ is empty. We consider a sampling time of $h = 0.03$ [s] in this example and synthesize a DT controller using  Theorem~\ref{THM:DissipativitySYnthesis} and the procedure indicated in Section~\ref{sec:special}, leading to \begin{equation}
K_\mathrm{d}(z) = \frac{431 z - 37.33}{z - 0.09988}.
\end{equation}
We compare this controller by a FOH and a Tustin discretization of $K_\mathrm{c}(s)$ with the same sampling time. In Fig.~\ref{fig:ControllerMatchingUnstableSys}, we compare the step response of the closed-loop system given by $\mathcal{G}$ and $\mathcal{K}_\mathrm{c}$ with the closed-loop formed by $\mathcal{G}$ and the tustin discretization of $\mathcal{K}_\mathrm{c}$, and the closed-loop systems formed by $\mathcal{G}$ and $\mathcal{K}_\mathrm{d}$, which results from the above redesign procedure. From this figure, we conclude that albeit some performance degradation is observed from the sampled-data controller the Tustin discretization fails to render the closed-loop systems stable, as the sampling time is chosen too large for the Tustin discretization to keep the closed-loop system stable. This demonstrates the need for having SD-controller-design methods, such as the ones presented in this paper.

\section{Conclusion}
In this paper, we have developed linear-matrix-inequality-based techniques for the design of sampled-data controllers that render the closed-loop system dissipative with respect to quadratic supply functions, which includes passivity and an upper-bound on the system's $\mathcal{H}_\infty$-norm as a special case. The results are applied to controller design for teleoperations that require passivity of the closed-loop system. It has been shown that using the sampled-data-synthesis techniques, a controller can be designed which renders the closed-loop system passive while sequential controller design and discretization does not render the closed-loop passive. Secondly, a filter-matching technique has been formulated, which minimizes the $\mathcal{H}_{\infty}$-norn and has been shown to produce a better approximation than traditional discretization techniques. Finally, the developed techniques have been shown to be applicable to the redesign of a given continuous-time controller as a sampled-data controller. The controller obtained through controller-matching is guaranteed to be stabilizing and has been shown to have better performance than the direct discretized version of the given continuous-time controller.

\balance
\bibliographystyle{agsm}
\bibliography{references}

\end{document}